\pdfoutput=1
\documentclass[11pt,twoside,a4paper,cmspaper,final,collab]{cms-tdr}
\def\svnVersion{76ef6d3-D}\def\svnDate{2021/03/01}\def\cmsCernNoTag{CERN-EP-2020-145}\def\cmsCernDate{\today}\def\cmsMessage{"Published in the European Physical Journal C as \href{http://dx.doi.org/10.1140/epjc/s10052-020-08817-8}{\doi{10.1140/epjc/s10052-020-08817-8}.}"}
\begin{document}\cmsNoteHeader{SMP-19-001}

\newcommand{\elelpr}{\ensuremath{2\ell'}\xspace}
\newcommand{\ptvecl}{\ensuremath{\ptvec^{\kern0.2em\ell}\xspace}}
\newcommand{\ZZ}{\ensuremath{\PZ\PZ}\xspace}
\newcommand{\ZZZ}{\ensuremath{\PZ\PZ\PZ}\xspace}
\newcommand{\pp}{\ensuremath{\Pp\Pp}\xspace}
\newcommand{\elel}{\ensuremath{2\ell}\xspace}
\newcommand{\twoeltwoel}{\ensuremath{\elel\elelpr}\xspace}
\newcommand{\elfourtau}{\ensuremath{2\ell2\ell^{\prime\prime}}\xspace}
\newcommand{\FxFx}{{\textsc{FxFx}}\xspace}
\newcommand{\matrixMC}{\textsc{Matrix}\xspace}
\newcommand{\cmsTable}[1]{\resizebox{\linewidth}{!}{#1}}
\newlength\cmsTabSkip\setlength{\cmsTabSkip}{1ex}
\ifthenelse{\boolean{cms@external}}{\providecommand{\cmsTabEnv}[1]{\resizebox{\linewidth}{!}{#1}}}{\providecommand{\cmsTabEnv}[1]{#1}}

\cmsNoteHeader{SMP-19-001}
\title{Measurements of \texorpdfstring{$\pp \to \ZZ$}{pp to ZZ} production cross sections and constraints on anomalous triple gauge couplings at \texorpdfstring{$\sqrt{s} = 13\TeV$}{sqrt(s) = 13 TeV}}
\titlerunning{$\pp \to \ZZ$ productions cross sections at $\sqrt{s}=13\TeV$}
\author{CMS Collaboration}

\date{\today}

\abstract{
The production of \PZ boson pairs in proton-proton (\pp) collisions, $\pp \to (\PZ/\gamma^*)(\PZ/\gamma^*) \to 2\ell 2\ell'$, where $\ell,\ell'$ = $\Pe$ or $\Pgm$, is studied at a center-of-mass energy of 13\TeV with the CMS detector at the CERN LHC. The data sample corresponds to an integrated luminosity of 137\fbinv, collected during 2016--2018.   
The \ZZ  production cross section,
$\sigma_{\text{tot}} ( \pp \to \ZZ ) = 17.4 \pm 0.3 \stat \pm 0.5 \syst \pm 0.4 \thy \pm 0.3  \lum\unit{pb}$,
measured for events with two pairs of opposite-sign, same-flavor leptons produced in the mass region $60 < m_{\ell^+\ell^-} < 120\GeV$ is consistent with standard model predictions.
Differential cross sections are
also measured and agree with theoretical  predictions.
The invariant mass distribution of the four-lepton system is used to set limits on
anomalous \ZZZ\ and $\ZZ\gamma$ couplings.
}

\hypersetup{%
pdfauthor={CMS Collaboration},%
pdftitle={Measurements of pp to ZZ production cross sections and constraints on anomalous triple gauge couplings at sqrt(s) = 13 TeV},%
pdfsubject={CMS},%
pdfkeywords={CMS, electroweak}}

\maketitle

\section{Introduction}

{\tolerance=800
Measurements of diboson production in proton-proton (\pp) collisions, such as \PZ boson pair (\ZZ) production, at the CERN LHC allow precision
tests of the standard model (SM).
In the SM, $\ZZ$ production proceeds mainly through quark-antiquark
$t$- and $u$-channel scattering
diagrams. In calculations at higher orders in quantum chromodynamics (QCD),
gluon-gluon fusion also contributes via
box diagrams with quark loops. 
There are no tree-level contributions to
$\ZZ$ production from triple gauge boson vertices in the SM.
Anomalous triple gauge couplings (aTGC)
$\ZZZ$ and $\ZZ\gamma$ are introduced using an effective Lagrangian
following Ref.~\cite{Hagiwara}.
In this parametrization, two $\ZZZ$ and two $\ZZ\gamma$ couplings are allowed by the 
electromagnetic gauge
invariance and Lorentz invariance for on-shell $\PZ$ bosons
and are parametrized by two
CP-violating ($f_4^{\PV}$) and two CP-conserving ($f_5^{\PV}$)
parameters,
where ${\PV} = (\PZ, \gamma)$.
Nonzero aTGC values could be induced by new physics models such as
supersymmetry~\cite{Gounaris:2000tb}.
The results
 can be also expressed in terms of parameters calculated 
within the effective field theory (EFT) framework, per convention used  in  Ref.~\cite{Degrande:2012wf} and references therein.  
In contrast to the anomalous couplings of electroweak (EW) vector bosons, the EFT
 framework allows an unambiguous calculation of loop effects and
provides a simpler interpretation of the results than the aTGC framework.
\par}

Previous measurements of the production cross section
for pairs of on-shell $\PZ$ bosons at the LHC
were performed
by the
CMS Collaboration
 with data sets corresponding to
integrated luminosities of 5.1\fbinv at $\sqrt{s} = 7\TeV$~\cite{Chatrchyan:2012sga}
and 19.6\fbinv at $\sqrt{s} = 8\TeV$~\cite{CMS:2014xja, Khachatryan:2015pba} in
the $\ZZ \to \elfourtau$ and $\ZZ \to \elel 2\nu$
decay channels, where $\ell = \Pe$ or $\Pgm$ and
$\ell^{\prime\prime} = \Pe, \Pgm$, or $\Pgt$; and with an integrated luminosity of
2.6\fbinv~\cite{Khachatryan:2016txa} and 35.9\fbinv~\cite{Sirunyan:2017zjc} at $\sqrt{s} = 13\TeV$ in
the $\ZZ \to \twoeltwoel$ decay channel, where
$\ell' = \Pe$ or $\Pgm$. All of the results agree with SM predictions.
The ATLAS Collaboration reported similar
results at $\sqrt{s} = 7$, 8, and 13\TeV~\cite{Aad:2012awa, Aad:2015rka, Aad:2015zqe, Aaboud:2019lxo, Aaboud:2019lgy, Aaboud:2017rwm},
which also agree with SM predictions.
These measurements are important to test predictions that were
recently made available at next-to-next-to-leading order (NNLO) in
QCD~\cite{Cascioli:2014yka,Grazzini:2015hta,Kallweit_2018}. A comparison of these predictions
with data for a range of center-of-mass energies provides an insight into the structure
of the EW gauge sector of the SM.

This paper reports a study of the \ZZ  production in the four-lepton decay channel
($\pp  \to  \twoeltwoel$, where $2\ell$ and $2\ell'$ indicate
pairs of opposite-sign electrons or muons) at $\sqrt{s} = 13\TeV$, with a data set corresponding to
an integrated luminosity of $137\fbinv$ recorded in 2016--2018.
Both $\PZ$ bosons are selected to be on-shell,
defined as the mass range 60--120\GeV.
Fiducial and total cross sections are measured, differential cross sections are presented as a function
of different kinematic variables.
The invariant mass distribution of the four-lepton system is used to search for
anomalous $\ZZZ$ and $\ZZ\gamma$ couplings.

\section{The CMS detector}

A detailed description
of the CMS detector, together with a definition of the coordinate system used
and the relevant kinematic variables, can be found
in Ref.~\cite{Chatrchyan:2008zzk}.

The central feature of the CMS apparatus is a superconducting solenoid of
6\unit{m} internal diameter, providing a magnetic field of 3.8\unit{T}.
Within the solenoid volume are a silicon pixel and strip
tracker, a lead tungstate crystal electromagnetic calorimeter (ECAL), and a
brass and scintillator hadron calorimeter,
which provide coverage in pseudorapidity $\abs{ \eta } < 1.479 $ in a cylindrical barrel
and $1.479 < \abs{ \eta } < 3.0$ in two endcap regions.
Forward calorimeters extend the
coverage provided by the barrel and
endcap detectors to $\abs {\eta} < 5.0$. Muons are detected in gas-ionization detectors embedded in
the steel flux-return yoke outside the solenoid
in the range $\abs{\eta} < 2.4$, with
detection planes made using three technologies: drift tubes, cathode strip
chambers, and resistive-plate chambers.

Electron momenta are estimated by combining energy measurements in the
ECAL with momentum measurements in the tracker. The momentum resolution for
electrons with transverse momentum $\pt \approx 45\GeV$
from $\PZ \to \Pep \Pem$ decays
ranges from 1.7\% for nonshowering electrons in the barrel region to 4.5\% for
showering electrons in the endcaps~\cite{Khachatryan:2015hwa}.
Matching muons to tracks identified in
the silicon tracker results in a $\pt$ resolution for
muons with $20 <\pt < 100\GeV$ of 1.3--2.0\% in the barrel and better than 6\%
in the endcaps. The \pt resolution in the barrel is better than 10\% for muons
with \pt up to 1\TeV~\cite{Chatrchyan:2012xi,Sirunyan:2018fpa}.

Events of interest are selected using a two-tiered trigger system~\cite{Khachatryan:2016bia}. The first level, composed of custom hardware processors, uses information from the calorimeters and muon detectors to select events at a rate of around 100\unit{kHz} within a time interval of less than 4\mus. The second level, known as the high-level trigger, consists of a farm of processors running a version of the full event reconstruction software optimized for fast processing, and reduces the event rate to around 1\unit{kHz} before data storage.

\section{Signal and background simulation}
\label{sec:mc}

{\tolerance=1600
Several Monte Carlo (MC) samples are used in the analysis to optimize the selection,
calculate the signal efficiency, and estimate background contamination.
The \PYTHIA~8.226 and 8.230~\cite{Sjostrand:2015,Alioli:2010xd} packages
are used for parton showering, hadronization, and
the underlying event simulation with the 
CUETP8M1 tune~\cite{Khachatryan:2015pea} and 
the parton distribution function (PDF)
NNPDF23\_lo\_as\_0130~\cite{nnpdf31} 
 for the 2016 data-taking period, and
the CP5 tune 
\cite{Sirunyan:2019dfx}
and the NNPDF 31\_nnlo\_as\_0118 PDF
for the 2017 and 2018 data-taking periods.
\par}

{\tolerance=800
Signal events are generated with
{\POWHEG}~2.0~\cite{Alioli:2008gx,Nason:2004rx,Frixione:2007vw,Alioli:2010xd,Melia:2011tj} at
next-to-leading order (NLO) in QCD for quark-antiquark ($\cPq\cPaq$) processes and
leading order (LO) for quark-gluon processes. This includes
$\ZZ$, $\PZ\gamma^\ast$, $\PZ$, $\gamma^\ast\gamma^\ast$
with a constraint of $m_{\ell\ell'} > 4\GeV$ applied to all pairs
of oppositely charged leptons at the generator level to avoid infrared divergences.
The gluon-gluon loop-induced process, $\Pg\Pg  \to \ZZ$, is simulated at LO
with \MCFM~v7.0~\cite{Campbell:2010ff}. It also includes interference with the SM Higgs off-shell
production. The SM Higgs decay is modeled with
\textsc{jhugen}~3.1.8~\cite{Gao:2010qx,Bolognesi:2012,Anderson:2013afp} at LO.  The cross sections are scaled to
correspond to cross section values calculated at NNLO in QCD for
$\cPq\cPaq  \to  \ZZ$~\cite{Cascioli:2014yka} (with a $K$~factor of 1.1) and at NLO in QCD for
$\Pg\Pg  \to  \ZZ$~\cite{Caola:2015psa} ($K$~factor of 1.7). 
Electroweak \ZZ  production in association with two jets is generated
with \MADGRAPH~\cite{mg_amcnlo} at LO. It amounts to approximately 1\% of the total number of \ZZ  events.
\par}

{\tolerance=800
Simulated events for the irreducible background processes containing four prompt leptons in the
final state, such as
$\ttbar\PZ$, 
$\PW\PW\PZ$, $\PW\PZ\PZ$, and $\PZ\PZ\PZ$, where the last three are combined and denoted as VVV, are generated with
\MGvATNLO v2.4.2~\cite{mg_amcnlo}
at NLO 
with zero or one outgoing partons in the matrix element calculation and merged with the
parton shower using the \FxFx scheme~\cite{Frederix:2012ps}. The same MC is used for $\PW\PZ$ simulation.
\par}

Event samples with aTGC contributions included are generated at LO with
\SHERPA~v2.1.1~\cite{Gleisberg:2008ta}.
The distributions from the \SHERPA samples are normalized such that the
total yield of the SM sample is the same as that of the {\POWHEG}+{\MCFM} sample.
More details are discussed in Section~\ref{sec:aTGC}.

The detector response is simulated using a detailed
description of the CMS detector implemented with the \GEANTfour
package~\cite{GEANT}. The reconstruction in simulation and data uses  
the same algorithms.
The simulated samples include additional interactions per bunch crossing,
referred to as pileup.
The simulated events are weighted so that the pileup distribution matches
the data.

{\tolerance=800 
Results are also compared to fixed-order predictions produced via the \matrixMC
framework~\cite{Grazzini:2017mhc}, a parton-level MC generator that uses tree and one-loop amplitudes 
from OpenLoops~2~\cite{Buccioni_2019} and two-loop amplitudes from Ref.~\cite{Gehrmann_2015},
capable of producing differential predictions 
at up to NNLO in QCD and NLO in EW, as implemented in \matrixMC~v2.0.0\_beta1~\cite{Grazzini_2020}.
The calculation is performed with the NNPDF31\_nnlo\_as\_0118\_luxqed~\cite{Bertone_2018} PDF set 
with dynamic renormalization ($\mu_{\mathrm{R}}$) and factorization scales ($\mu_{\mathrm{F}}$) set to the 
four lepton mass for the differential and fiducial predictions, and with fixed scale set to the nominal \PZ boson mass 
for the total cross section. 
The quark-induced processes are calculated at NNLO in QCD and NLO in EW. 
The gluon-induced contribution is calculated at NLO in QCD~\cite{Grazzini_2019}. 
Photon-induced contributions 
are also included at up to NLO EW. The calculation uses massless leptons, which leads to a divergence 
at low dilepton mass. To avoid this divergence, we impose the 
requirement $\pt^{\ell} > 5\GeV$ on the photon-induced component for total cross section predictions. 
With this condition, the photon-induced contribution is less than 1\% of the total production rate. 
The quark-induced NNLO QCD and NLO EW contributions are combined multiplicatively, and the gluon-
and photon-induced contributions are combined additively following the procedure described in Ref.~\cite{Grazzini_2020}.
The predictions reported here are consistent with those published in Refs.~\cite{Cascioli:2014yka,Grazzini:2015hta,Kallweit_2018}.
\par}

\section{Event reconstruction}
\label{sec:eventreconstruction}

{\tolerance=800
Individual particles---electrons, muons, photons, charged and neutral
hadrons---in each collision event are identified and reconstructed
with the CMS particle-flow (PF) algorithm~\cite{Sirunyan:2017ulk}
from a combination of signals from all subdetectors.
Reconstructed electrons~\cite{Khachatryan:2015hwa} and muons~\cite{Chatrchyan:2012xi}
are considered as lepton candidates 
if they have $\pt^{\Pe} > 7\GeV$ and $\abs{\eta^{\Pe}} < 2.5$ or
$\pt^{\Pgm} > 5\GeV$ and $\abs{\eta^{\Pgm}} < 2.4$.
\par}

{\tolerance=800
Lepton candidates are also required to originate from the primary vertex, defined as the
reconstructed \pp interaction vertex with the largest value of
summed physics object $\pt^2$. The physics objects used in the primary vertex definition
are the objects returned by a jet-finding
algorithm~\cite{Cacciari:2008gp,Cacciari:2011ma} applied to all charged tracks
associated with the vertex.
The distance of closest approach between
each lepton track and the primary vertex
is required to be less than 0.5\unit{cm} in the plane transverse to the beam axis,
and less than 1\unit{cm} in the direction along the beam axis.
Furthermore, the significance of the three-dimensional impact parameter relative
to the primary vertex, $\mathrm{SIP_{3D}}$, is required to satisfy
$\mathrm{SIP_{3D}} \equiv \abs{ \mathrm{IP} / \sigma_\mathrm{IP}} < 4$
for each lepton, where $\mathrm{IP}$ is the distance
of closest approach of each lepton track to the primary vertex
and $\sigma_\mathrm{IP}$ is its associated uncertainty.
\par}

Lepton candidates are required to be isolated from other particles in the event. The
relative isolation is defined as
\begin{linenomath}
\ifthenelse{\boolean{cms@external}}{
\begin{multline}
        R_\text{iso} = \bigg[ \sum_{\substack{\text{charged} \\ \text{hadrons}}} \!\! \pt \,  \\
                             +\max\bigg(0, \sum_{\substack{\text{neutral} \\ \text{hadrons}}} \!\! \pt
                                       + \, \sum_{\text{photons}} \!\! \pt \, - \, \pt^\mathrm{PU}
                                       \bigg)\bigg] \bigg/ \pt^{\ell},
        \label{eq:iso}
\end{multline}
}{
\begin{equation}
        R_\text{iso} = \bigg[ \sum_{\substack{\text{charged} \\ \text{hadrons}}} \!\! \pt \, + \,
                             \max\big(0, \sum_{\substack{\text{neutral} \\ \text{hadrons}}} \!\! \pt
                                       \, + \, \sum_{\text{photons}} \!\! \pt \, - \, \pt^\mathrm{PU}
                                       \big)\bigg] \bigg/ \pt^{\ell},
        \label{eq:iso}
\end{equation}
}
\end{linenomath}
where the sums run over the charged and neutral hadrons, and photons identified
by the PF algorithm, in a cone defined by
$\Delta R \equiv \sqrt{\smash[b]{\left(\Delta\eta\right)^2 + \left(\Delta\varphi\right)^2}} < 0.3$
around the lepton momentum direction, where $\varphi$ is the azimuthal angle in radians.
To minimize the contribution of charged particles from pileup to the isolation calculation,
charged hadrons are included only if they originate from the
primary vertex. The contribution of
neutral particles from pileup is $\pt^\mathrm{PU}$. For electrons, $\pt^\mathrm{PU}$
is evaluated with the ``jet area'' method described in Ref.~\cite{Cacciari:2007fd};
for muons, it is 
 half of the summed  $\pt$ of all charged particles in the cone originating
from pileup vertices. The average  factor of
one half accounts for the expected ratio of neutral to charged particle production 
in hadronic interactions. A lepton is considered isolated if
$R_\text{iso} < 0.35$.

The lepton reconstruction, identification, and isolation efficiencies
are measured with a ``tag-and-probe''
technique~\cite{CMS:2011aa} applied to a sample of $\PZ  \to  \ell^+\ell^-$ data events.
The measurements are performed in several bins of $\pt^{\ell} $ and $\abs{\eta^\ell}$.
The electron reconstruction and selection efficiency in the ECAL barrel (endcaps) varies from
about 85 (77)\% at $\PT^{\Pe} \approx 10\GeV$
to about 95 (89)\% for $\PT^{\Pe} \geq 20\GeV$,
whereas in the barrel-endcap transition region this efficiency is about 85\% averaged
over electrons with $\pt^{\Pe} > 7\GeV$.
The muons are reconstructed and identified with efficiencies above ${\sim}98\%$
within $\abs{\eta^{\Pgm}} < 2.4$.

\section{Event selection}

The primary triggers for this analysis require the presence of a
pair of loosely isolated leptons of the same or different
flavors~\cite{Khachatryan:2016bia}.
The highest \pt lepton must have $\pt^{\ell} > 17\GeV$, and the
subleading lepton must have
$\pt^{\Pe} > 12\GeV$ if it is an electron or $\pt^{\Pgm} > 8\GeV$
if it is a muon. The tracks of the triggering leptons are required to originate within
2~mm of each other in the plane transverse to the beam axis. Triggers
requiring a triplet of lower-\pt leptons
with no isolation criteria, or a single high-\pt electron or muon, are also used.
An event is accepted if it passes any trigger regardless of the decay channel.
The total trigger efficiency for events within the acceptance of this
analysis is greater than 98\%.

The four-lepton candidate selection is based on the one used in the recent CMS Higgs boson measurement 
~\cite{Sirunyan:2017exp}. A signal event must contain at least two
$\PZ/\gamma^{\ast}$ candidates, each formed from an oppositely charged
pair of isolated electron or muon candidates.
Among the four leptons, the highest \pt lepton must have $\pt > 20\GeV$, and
the second-highest \pt lepton must have $\pt^{\Pe} > 12\GeV$ if it is an electron
or $\pt^{\Pgm} > 10\GeV$ if it is a muon.
All leptons are required to be separated from each other by
$\Delta R \left(\ell_1, \ell_2 \right) > 0.02$,
and electrons are required to be separated from muons by
$\Delta R \left(\Pe, \mu \right) > 0.05$.

Within each event, all
permutations of leptons giving a valid pair of $\PZ/\gamma^{\ast}$
candidates are considered separately.
Within each four-lepton candidate, the
dilepton candidate with an invariant mass closest to 91.2\GeV, taken as
the nominal $\PZ$ boson mass~\cite{Tanabashi:2018oca}, is denoted $\PZ_1$ and is required to have a
mass greater than 40\GeV. The other dilepton candidate is denoted $\PZ_2$ and is
required to have a mass greater than 4\GeV.
Both $m_{Z_1}$ and $m_{Z_2}$ are required to be less than 120\GeV.
All pairs of oppositely charged leptons in the four-lepton candidate are required to
have $m_{\ell \ell'} > 4\GeV$ regardless of their flavor.
In the rare case
of further ambiguity, which occurs in less than 0.5\% of events when
five or more passing lepton candidates are found, the $\PZ_2$ candidate
that maximizes the scalar $\pt$ sum of the four leptons is chosen.

The $\pp  \to  \ZZ$ cross section is measured using events where both
$m_{\PZ_1}$ and $m_{\PZ_2}$ are greater than 60\GeV.
Decays of the $\PZ$ bosons to
$\PGt$ leptons with subsequent decays to electrons and muons are heavily
suppressed by the requirements on lepton $\pt$, and the contribution of such
events is less than 0.5\% of the total $\ZZ$ yield. If these events pass the
selection requirements of the analysis, they are considered signal, although they
are not considered at generator level in the cross section measurement procedure.
Thus, the correction for possible $\tau$ decays is included in the efficiency
calculation.

\section{Background estimation}

{\tolerance=800
The requirement of four well-reconstructed and isolated lepton candidates strongly suppresses any
background; therefore this analysis has very low background contributions, dominated by
$\PZ$ boson and $\PW\PZ$ diboson production in association with
jets, and by \ttbar production.
In a small fraction of cases, particles from jet fragmentation satisfy both lepton identification and
isolation criteria, and thus are misidentified as signal leptons. This background is estimated using control data samples, as decribed below.
\par}

The probability for such objects to be selected
is measured from a sample of
$\PZ$+$\ell_\text{candidate}$ events, where $\PZ$ denotes a pair of
oppositely charged, same-flavor leptons that pass all analysis requirements and
satisfy $\abs{ m_{\ell^+\ell^-} - m_{\PZ}} < 10\GeV$, where
$m_{\PZ}$ is the nominal $\PZ$ boson mass. Each event in this sample must have exactly one
additional object $\ell_\text{candidate}$ that passes relaxed identification requirements with
no isolation requirements applied. The misidentification probability for
each lepton flavor, measured in bins of lepton candidate $\pt$ and $\eta$,
is defined as the ratio between the number of candidates that pass the final
isolation and identification requirements and the total number of candidates in the sample.
The number of $\PZ$+$\ell_\text{candidate}$ events is corrected for the contamination
from $\PW\PZ$ production and for $\ZZ$ events in which one lepton is not
reconstructed. These events have a third genuine, isolated lepton that must be excluded
from the misidentification probability calculation. The $\PW\PZ$ contamination is suppressed by requiring the missing
transverse momentum $\ptmiss$ to be below 25\GeV. The $\ptmiss$ is defined
as the magnitude of the missing transverse momentum vector $\ptvecmiss$,
the projection onto the plane transverse to the beams of the negative
vector sum of the momenta of all reconstructed PF candidates in the event,
corrected for the jet energy scale.
The transverse mass, calculated as
$\mT \equiv \sqrt{\smash[b]{(\pt^\ell + \ptmiss )^2 -
(\ptvecl + \ptvecmiss )^2}}$, is required to be less than 30\GeV. The residual contribution
of $\PW\PZ$ and $\ZZ$ events, which can be up to a few percent of the
events with $\ell_\text{candidate}$ passing all selection criteria, is estimated
from simulation and subtracted.

To account for all sources of background events, two control samples are
used to estimate the number of background events
in the signal regions. Both are defined as samples that contain events with
a dilepton candidate satisfying all requirements ($\PZ_1$) and
two additional lepton candidates $\ell^{+}\ell^{-}$.
In one control sample, enriched in $\PW\PZ$ events, one
$\ell$ candidate is required to satisfy the full
identification and isolation criteria and the other must fail the full criteria
and instead satisfy only the relaxed ones; in the other, enriched in
$\PZ$+jets events, both $\ell$
candidates must satisfy the relaxed criteria, but fail the full criteria.
The additional leptons must have opposite charges and the same
flavor ($\Pe^{\pm}\Pe^{\mp}$ and $\Pgm^{\pm}\Pgm^{\mp}$).
From this set of events, the expected number of background events in the
signal region, denoted ``$\PZ$+X'' in the figures, is obtained
by scaling the number of observed $\PZ_1+\ell^{+}\ell^{-}$ events
by the misidentification probability for each lepton failing the selection.
The procedure is described in more detail in Ref.~\cite{Chatrchyan:2013mxa}.

In addition to this reducible background, which contributes to 
approximately 1--2\% of the \ZZ events, 
the $\ttbar\PZ$ and VVV processes 
with four prompt leptons are estimated from simulated samples
to be around 1--1.5\% of the expected $\ZZ \to 2\ell2\ell'$ yield.
The total background contributions to the 
$\ZZ \to 2\ell2\ell'$ signal regions are summarized in
Section~\ref{sec:xsec}.

\section{Systematic uncertainties}\label{sec:systematics}

The major sources of systematic uncertainty and their effect on the
measured cross sections are summarized in Table~\ref{table:systematics}.
The lepton identification, isolation, and track reconstruction efficiencies in simulation are corrected
with scaling factors derived with a tag-and-probe method and applied as a
function of lepton $\pt$ and $\eta$.
To estimate the uncertainties associated with the tag-and-probe technique, the total yield is
recomputed with the scaling factors varied up and down by the tag-and-probe
fit uncertainties. The uncertainties associated with the lepton efficiency in
the $\ZZ  \to  2\ell2\ell'$ 
signal regions are 5\% in the $4\Pe$, 3\%
in the  $2\Pe 2\mu$, and 2\% in the $4\mu$ final states.

In both data and simulated event samples, trigger efficiencies are evaluated with
a tag-and-probe technique. The ratio of the trigger efficiency estimated using data to the one estimated
with simulation is applied
to simulated events, and the size of the resulting change in the expected yield is
taken as the uncertainty in the determination of the trigger efficiency.
This uncertainty is
around 1--2\% of the final estimated yield. 

\begin{table}[htb]
\centering
\topcaption{
  The contributions of each source of systematic uncertainty in the
  cross section measurements. The integrated luminosity uncertainty, and the PDF and scale
  uncertainties, are considered separately. All other uncertainties are added
  in quadrature into a single systematic uncertainty. Uncertainties that vary by
  decay channel are listed as ranges.
}
\begin{tabular}{lc}
\hline
Uncertainty & Range of values  \\
\hline
Lepton efficiency         & 2--5\%       \\
Trigger efficiency        & 1--2\%        \\
Background                & 0.6--1.3\%    \\
Pileup                    & 1\%        \\
$\mu_\mathrm{R}$, $\mu_\mathrm{F}$          & 1\%    \\
PDF                       & 1\%           \\
NNLO/NLO corrections       & 1\%           \\
Integrated luminosity     & 2.5\% (2016), 2.3\% (2017),  \\
     & 2.5\% (2018)   \\
\hline
\end{tabular}
\label{table:systematics}
\end{table}

The largest uncertainty in the estimated background yield arises from
differences in sample composition between the $\PZ$+$\ell_\text{candidate}$ control sample
used to calculate the lepton misidentification probability and the
$\PZ+\ell^+\ell^-$ control sample. An additional uncertainty arises
from the limited number of events in the $\PZ$+$\ell_\text{candidate}$ sample. A
systematic uncertainty of 40\% is applied to the lepton misidentification probability
to cover both effects. Its impact varies by channel, but
is of the order of 1\% of the total expected yield.

The modeling of pileup relies on the total inelastic \pp cross section~\cite{Sirunyan:2018nqx}.
The pileup uncertainty
is evaluated by varying this cross section up and down by 5\%.

Uncertainties because of factorization ($\mu_\mathrm{F}$) and
renormalization ($\mu_\mathrm{R}$) scale choices on the $\ZZ \to 2\ell2\ell'$
acceptance are evaluated with {\POWHEG}+\MCFM by varying $\mu_\mathrm{F}$ and $\mu_\mathrm{R}$
up and down by a factor of two with respect to the default values
$\mu_\mathrm{F} = \mu_\mathrm{R} = m_{\ZZ}$, where $m_{\ZZ}$ is the invariant
mass of the \ZZ  system. All combinations are considered
except those in which $\mu_\mathrm{F}$ and $\mu_\mathrm{R}$ differ by a factor of
four. Parametric uncertainties
(PDF+$\alpS$) are evaluated according to the PDF4LHC
prescription~\cite{Butterworth:2015oua} in the acceptance calculation,
and with NNPDF3.0~\cite{nnpdf} in the cross section calculations.
An additional theoretical uncertainty arises from scaling the {\POWHEG}
$\cPq\cPaq  \to  \ZZ$ simulated sample from its NLO cross section to the
NNLO prediction, and the \MCFM $\Pg\Pg  \to  \ZZ$ samples
from their LO cross sections to the NLO predictions. The change in the
acceptance corresponding to this scaling procedure is about 1\%.

The uncertainty in the integrated luminosity of the data samples
is 2.5\% (2016)~\cite{CMS-PAS-LUM-17-001}, 2.3\% (2017)~\cite{CMS-PAS-LUM-17-004}, and 2.5\% (2018)~\cite{CMS-PAS-LUM-18-002}.
Since the luminosity uncertainty contains a significant uncorrelated portion, the 
relative luminosity uncertainty of the whole sample is smaller than for each individual year.

The same uncertainties are valid for both total and differential cross section measurements,
but for the differential one there is also an additional uncertainty related to 
the unfolding procedure described in Section~\ref{sec:differentials}. It is estimated using
\MGvATNLO instead of {\POWHEG}+\MCFM in unfolding. The unfolding uncertainty is included
 in the results and plots together with other uncertainites, 
but its effect is small compared to the statistical uncertainties of the measurement.

\section{Cross section measurement}
\label{sec:xsec}

\begin{figure*}[htbp]
\begin{center}
{\includegraphics[width=0.42\textwidth]{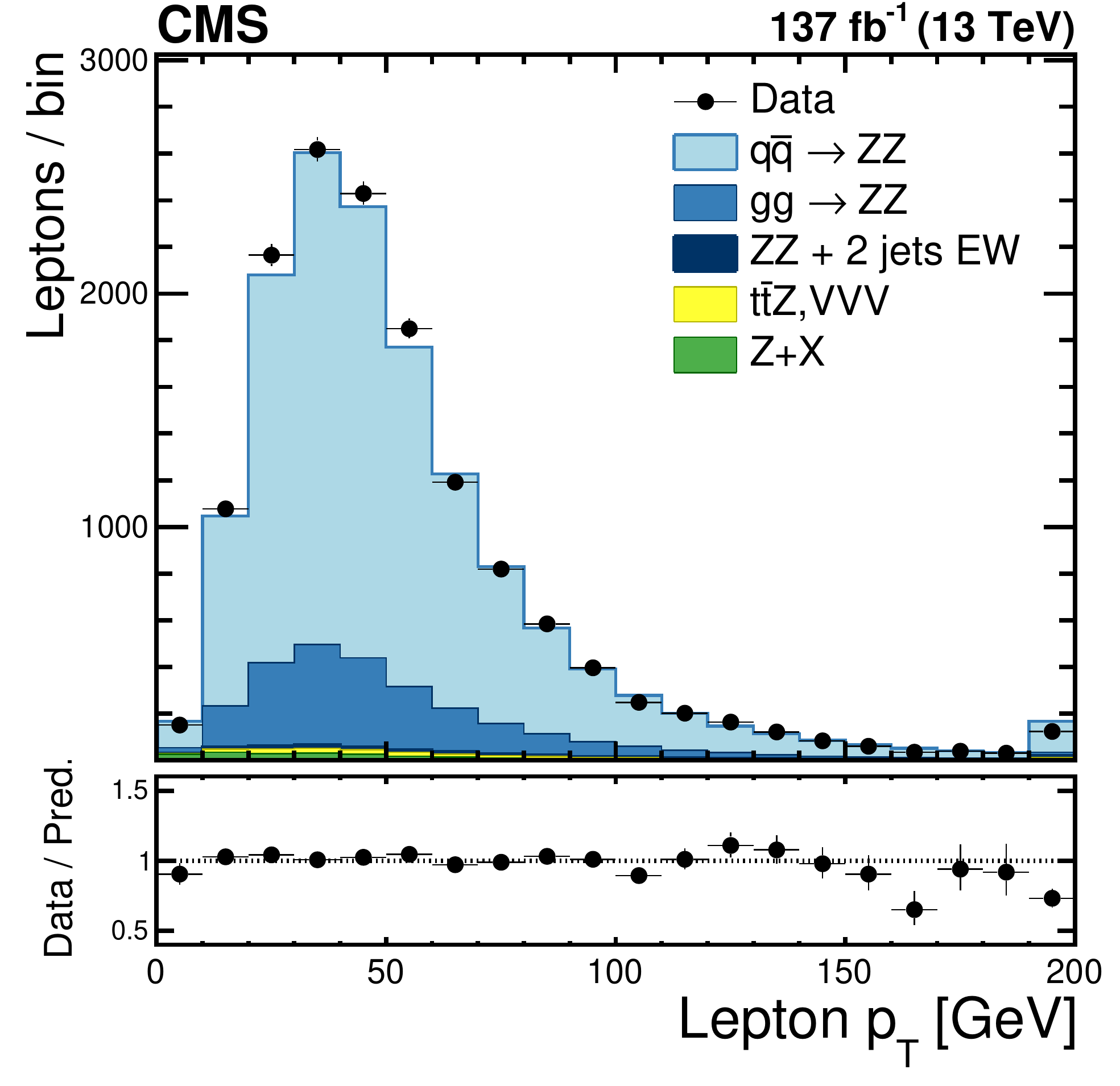}}
{\includegraphics[width=0.42\textwidth]{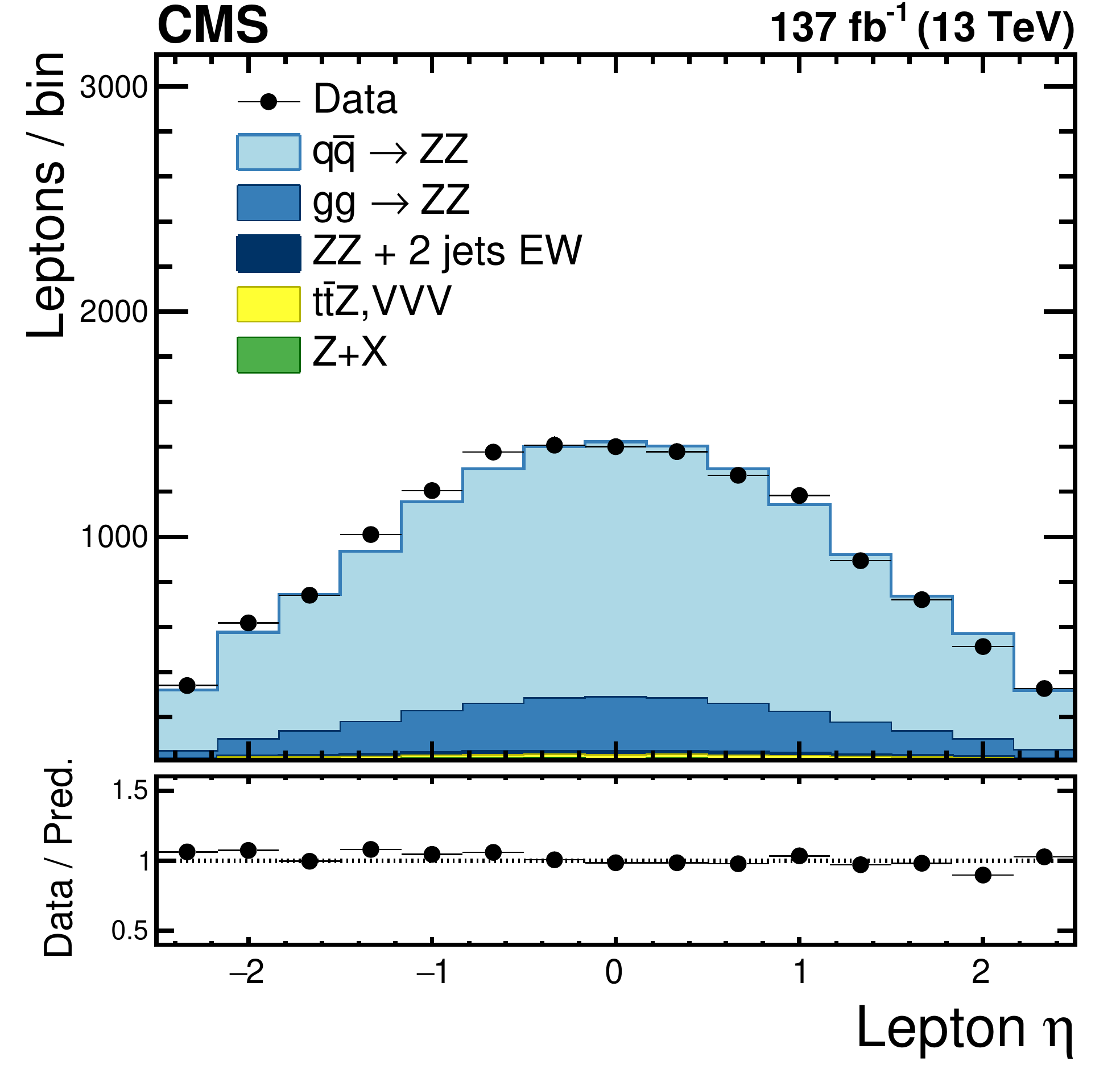}}
\caption{
Distributions of (left) transverse momentum and (right) pseudorapidity for
individual leptons.
Points represent the data with error bars showing the statistical uncertainties, histograms 
the expected SM predictions and reducible background estimated from
data.
The \pt distributions includes overflow in the last bin.
}
\label{fig:results_smp_zPt}
\end{center}
\end{figure*}

\begin{figure*}[htbp]
\begin{center}
{\includegraphics[width=0.42\textwidth]{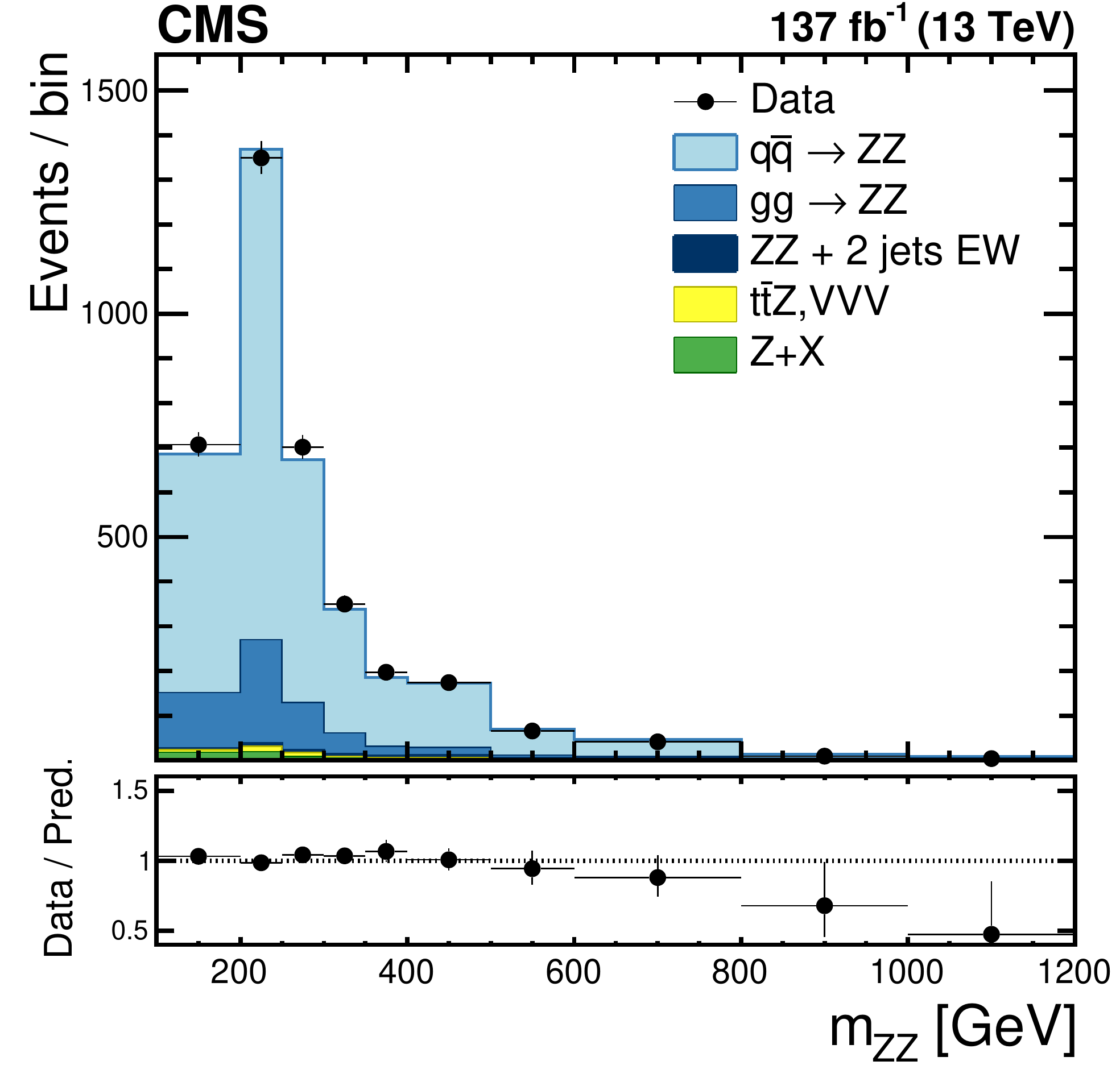}}
{\includegraphics[width=0.42\textwidth]{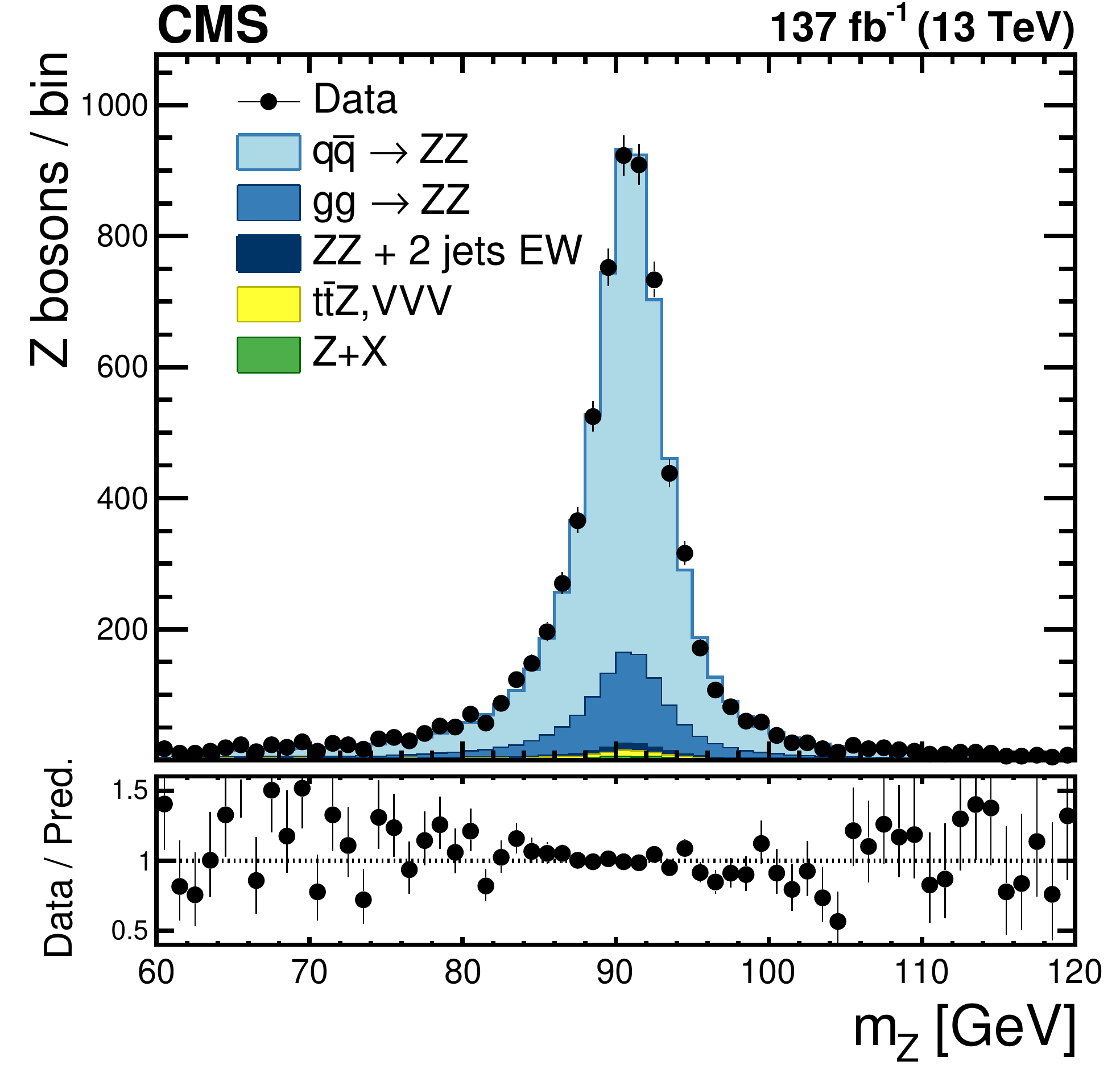}} \\
{\includegraphics[width=0.42\textwidth]{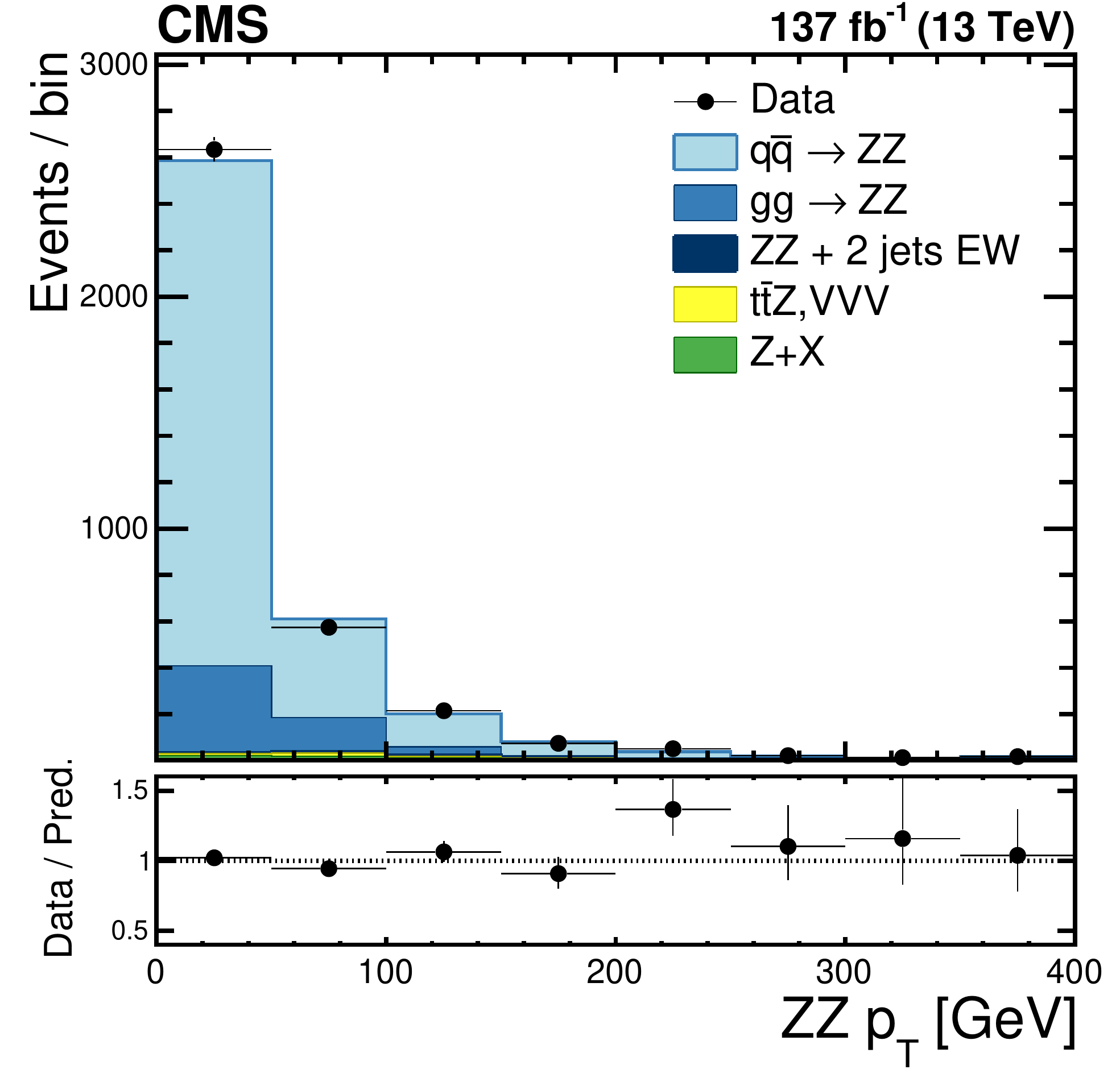}}
{\includegraphics[width=0.42\textwidth]{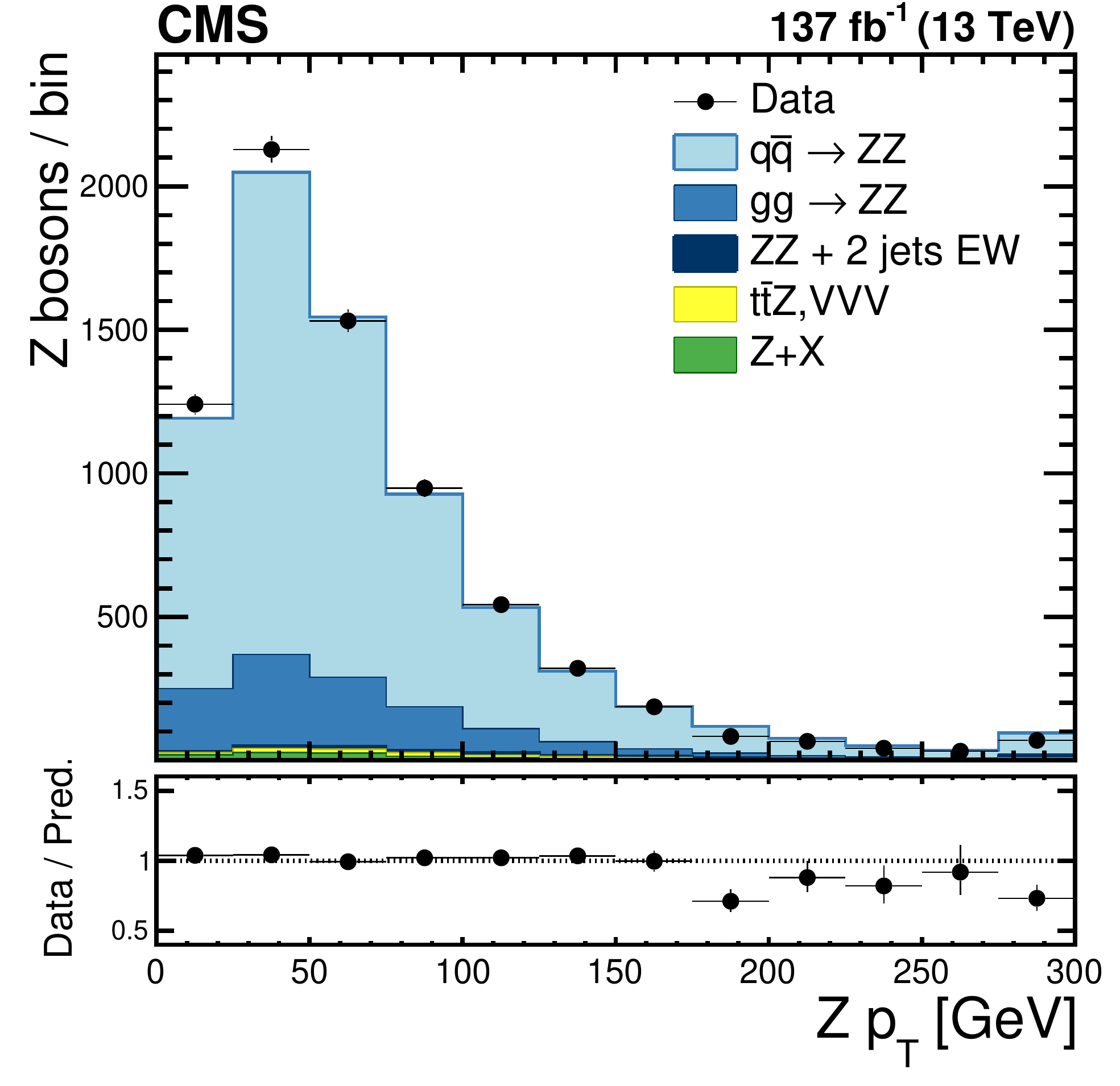}}
\caption{
Distributions of (upper left) $m_{\ZZ}$ for \ZZ  events
with $60 < m_{\PZ_1, \PZ_2} < 120 \GeV$;
(upper right) mass of selected \PZ boson candidates; (lower left) transverse momentum of the $\ZZ$ system;
(lower right) transverse momentum of individual \PZ boson candidates.
Points represent the data with error bars showing the statistical uncertainties, histograms 
the expected SM predictions and reducible background estimated from
data. All \pt and $m_{\ZZ}$ distributions include overflow in the last bin.
\label{fig:results_smp_Mass}
}
\end{center}
\end{figure*}

{\tolerance=800
The \pt and $\eta$ distributions for
individual leptons are shown in Fig.~\ref{fig:results_smp_zPt}.
Both distributions contain four leptons per event.
The invariant mass of the \ZZ  system, the individual mass of  
reconstructed \PZ boson candidates
in the \ZZ  events, and their corresponding \pt distributions are shown in 
Fig.~\ref{fig:results_smp_Mass}. 
The last bins in $m_{\ZZ}$ and all \pt distributions contain events from the overflow.
The $m_\PZ$ and \PZ \pt distributions contain two \PZ candidates for each event. 
These distributions are shown for data and simulated events to demonstrate comparisons
with SM expectations. The signal expectations include contributions from $\ZZ$
production shown separately for
$\cPq\cPaq  \to  \ZZ$, 
$\Pg\Pg  \to  \ZZ$, and EW \ZZ processes in all figures and combined as ``Signal'' in Table~\ref{table:results_smp}. 
The EW \ZZ  production 
contributes to approximately  1\% of the total number of \ZZ  events. 
\par}

{\tolerance=800
The irreducible
background, 
which amounts to 1--1.5\% of the total \ZZ  yield, and
reducible background are combined  as ``Background'' in  Table~\ref{table:results_smp}.
The total background in this analysis is $\approx 3\%$.
The estimated yields agree well with the measured ones. 
The individual distributions are well described, except the $m_{\ZZ}$ distribution at high values of invariant masses and the $\pt^\PZ$ distribution at high values of $\pt$.
These are regions where the EW corrections may become important and will be discussed later in 
Section~\ref{sec:aTGC}. 
\par}

\begin{table*}[htbp]
\centering
\topcaption{ Observed and expected prefit yields of $\ZZ$ events,
and estimated yields of background
events, shown for each final state and
combined. The statistical (first) and systematic (second)  uncertainties are presented.
}
\cmsTable{
\begin{tabular}{lcccc}
\hline
Process       & $\Pe\Pe\Pe\Pe$  & $\Pe\Pe\Pgm\Pgm$  & $\Pgm\Pgm\Pgm\Pgm$ & $2\ell2\ell'$ \\
\hline
& \multicolumn{4}{c}{2016} \\
 Background & 6.7 $\pm$ 0.6 $\pm$ 1.8 & 11.4 $\pm$ 0.8 $\pm$ 1.9 & 5.5 $\pm$ 0.5 $\pm$ 0.9 & 23.6 $\pm$ 1.1 $\pm$ 4.1 \\ 
 Signal & 167.7 $\pm$ 1.0 $\pm$ 10.0 & 434.2 $\pm$ 1.6 $\pm$ 17.3 & 273.3 $\pm$ 1.3 $\pm$ 8.2 & 875.2 $\pm$ 2.3 $\pm$ 31.1 \\[\cmsTabSkip] 
 Total expected & 174.4 $\pm$ 1.2 $\pm$ 10.4 & 445.6 $\pm$ 1.8 $\pm$ 17.7 & 278.8 $\pm$ 1.4 $\pm$ 8.4 & 898.8 $\pm$ 2.6 $\pm$ 32.0 \\ 
Data      &          176              &            478             &            296             &            950      \\[\cmsTabSkip]
& \multicolumn{4}{c}{2017} \\
 Background & 6.3 $\pm$ 0.4 $\pm$ 1.5 & 12.1 $\pm$ 0.8 $\pm$ 1.8 & 7.9 $\pm$ 0.7 $\pm$ 1.6 & 26.3 $\pm$ 1.2 $\pm$ 4.5 \\ 
 Signal & 200.8 $\pm$ 0.3 $\pm$ 12.0 & 511.7 $\pm$ 0.6 $\pm$ 20.4 & 322.5 $\pm$ 0.5 $\pm$ 9.6 & 1035.0 $\pm$ 0.8 $\pm$ 36.9 \\[\cmsTabSkip] 
 Total expected & 207.1 $\pm$ 0.6 $\pm$ 12.4 & 523.8 $\pm$ 1.0 $\pm$ 20.9 & 330.4 $\pm$ 0.8 $\pm$ 9.9 & 1061.3 $\pm$ 1.4 $\pm$ 38.0 \\ 
Data      &      193                   &       540                   &       328                   &    1061            \\[\cmsTabSkip]
& \multicolumn{4}{c}{2018} \\
 Background & 9.9 $\pm$ 0.6 $\pm$ 2.5 & 23.2 $\pm$ 1.1 $\pm$ 4.2 & 15.6 $\pm$ 1.1 $\pm$ 4.0 & 48.7 $\pm$ 1.7 $\pm$ 9.7 \\ 
 Signal & 305.2 $\pm$ 0.4 $\pm$ 18.2 & 758.5 $\pm$ 0.8 $\pm$ 30.1 & 467.3 $\pm$ 0.6 $\pm$ 13.9 & 1531.0 $\pm$ 1.0 $\pm$ 54.7 \\[\cmsTabSkip] 
 Total expected & 315.1 $\pm$ 0.8 $\pm$ 18.7 & 781.7 $\pm$ 1.4 $\pm$ 31.1 & 482.9 $\pm$ 1.3 $\pm$ 14.8 & 1579.7 $\pm$ 2.0 $\pm$ 56.6 \\ 
Data      &      309                   &      797                    &     480                     &       1586         \\
\hline
\end{tabular}
}
\label{table:results_smp}
\end{table*}

The measured yields are used to evaluate the $\ZZ$ 
production cross section in the fiducial phase space.
The signal acceptance is evaluated from simulation and corrected
for each individual lepton flavor in bins of $\pt$ and $\eta$
using factors obtained with
the  tag-and-probe technique.
To include all final states in the cross section calculation,
a simultaneous fit to the number of  observed events in all decay channels
is performed. The likelihood is composed as a combination of
individual channel likelihoods for
the signal and background hypotheses
with the statistical
and systematic uncertainties treated as scaling nuisance parameters.
The combination of various data-taking periods is performed treating the theoretical 
uncertainties as fully correlated
among various periods, whereas the experimental uncertainties
are either correlated or uncorrelated, depending on their origin.

The fiducial phase space for the
$\ZZ \to 2\ell2\ell'$
cross section measurement is defined as:
$\pt^{\ell_1} > 20\GeV$,
$\pt^{\ell_2} > 10\GeV$,
$\pt^{\ell_{3,4}} > 5\GeV$, 
$\abs{\eta^{\ell}} < 2.5$,
$m_{2\ell} > 4\GeV$ (any opposite-sign same-flavor pair),
$60 < m_{\PZ_1}, m_{\PZ_2} < 120\GeV$.
The generator-level leptons used for the fiducial cross section calculation
are ``dressed'' by adding the momenta of generator-level photons within
$\Delta R\left(\ell,\gamma\right) < 0.1$ from the lepton momenta directions.

\begin{table}[htbp]
    \centering
  \topcaption{ Measured fiducial cross section for each data sample and combined. The first uncertainty is statistical, the second is experimental systematic, and the third is associated with the integrated luminosity.}
\label{tab:fiducial}
\begin{tabular}{ lc }
\hline
      Year       & Fiducial cross section, fb  \\
\hline
   2016       & $ 42.0 \pm 1.4 \stat \pm 1.3 \syst _{-1.0}^{+1.1} \lum $    \\
   2017       & $ 39.6 \pm 1.2 \stat _{-1.2}^{+1.3} \syst _{-0.9}^{+1.0} \lum $    \\
   2018       & $ 39.7 \pm 1.0 \stat _{-1.1}^{+1.3} \syst \pm 1.0 \lum $    \\[\cmsTabSkip]
  Combined    & $ 40.5 \pm 0.7 \stat \pm 1.1 \syst \pm 0.7 \lum $    \\
\hline
\end{tabular}
\end{table}

\begin{table}[htbp]
    \centering
  \topcaption{ Measured total $\sigma (\Pp\Pp \to \ZZ)$ cross section for each data sample and combined. The first uncertainty is statistical, the second is experimental systematic, the third is theoretical systematic. The fourth uncertainty is associated with the integrated luminosity.}
\label{tab:total}
\cmsTabEnv{
\begin{tabular}{ lc }
\hline
      Year       & Total cross section, pb  \\
\hline
   2016       & $ 18.1 \pm 0.6 \stat _{-0.5}^{+0.6} \syst \pm 0.4 \thy _{-0.4}^{+0.5} \lum $    \\
   2017       & $ 17.0 \pm 0.5 \stat _{-0.5}^{+0.6} \syst \pm 0.4 \thy \pm 0.4 \lum $    \\
   2018       & $ 17.1 \pm 0.4 \stat \pm 0.5 \syst \pm 0.4 \thy \pm 0.4 \lum $    \\[\cmsTabSkip]
  Combined    & $ 17.4 \pm 0.3 \stat \pm 0.5 \syst \pm 0.4 \thy \pm 0.3 \lum $    \\
\hline
\end{tabular}
}
\end{table}

The measured $\ZZ$ fiducial cross section presented in Table~\ref{tab:fiducial} can be compared to
$39.3^{+0.8}_{-0.7} \pm 0.6\unit{fb}$ calculated with {\POWHEG}+\MCFM
using the same settings as the simulated samples with $K$~factors applied.
The first uncertainty corresponds to the factorization and renormalization 
scales and the second to  
 PDF, as described above. The {\POWHEG} 
calculations used dynamic factorization and renormalization scales
$\mu_\mathrm{F} = \mu_\mathrm{R} = m_{2\ell2\ell'}$, whereas the
contribution from \MCFM is computed with dynamic scales
$\mu_\mathrm{F} = \mu_\mathrm{R} = 0.5 m_{2\ell2\ell'}$.
It can also be compared to the prediction from \matrixMC~v2.0.0\_beta1 of $38.0^{+1.1}_{-1.0}$.
The uncertainty in the \matrixMC prediction includes only the uncertainty due to the 
variation of $\mu_\mathrm{F}$ and $\mu_\mathrm{R}$.

{\tolerance=800
The total $\ZZ$ production cross section for both dileptons produced in the
mass range 60--120\GeV and $m_{\ell^+\ell^{\prime -}} > 4\GeV$ is 
presented in Table~\ref{tab:total}. 
The nominal branching fraction
$\mathcal{B}(\cPZ \to \ell^+\ell^-) = 0.03366$
 is used~\cite{Tanabashi:2018oca}.
The measured total cross section can be compared to the theoretical value of
$16.9^{+0.6}_{-0.5} \pm 0.2\unit{pb}$, calculated
from {\POWHEG}+\MCFM with the same settings that is used for
$\sigma_{\mathrm{fid}} (\pp \to \ZZ \to 2\ell2\ell')$.
It can also be compared to
$16.5^{+0.6}_{-0.5}$\unit{pb}, 
calculated with
\matrixMC~v2.0.0\_beta1, or
$15.0^{+0.7}_{-0.6} \pm 0.2$\unit{pb}, calculated with \MCFM at NLO in QCD with
additional contributions from LO $\Pg\Pg  \to  \ZZ$ diagrams and
with the NLO NNPDF3.0 PDF set and fixed factorization
and renormalization 
scales set to $\mu_\mathrm{F} = \mu_\mathrm{R} = m_{\PZ}$.
\par}

\begin{figure}[htbp]
\centering
\includegraphics[width=0.95\linewidth]{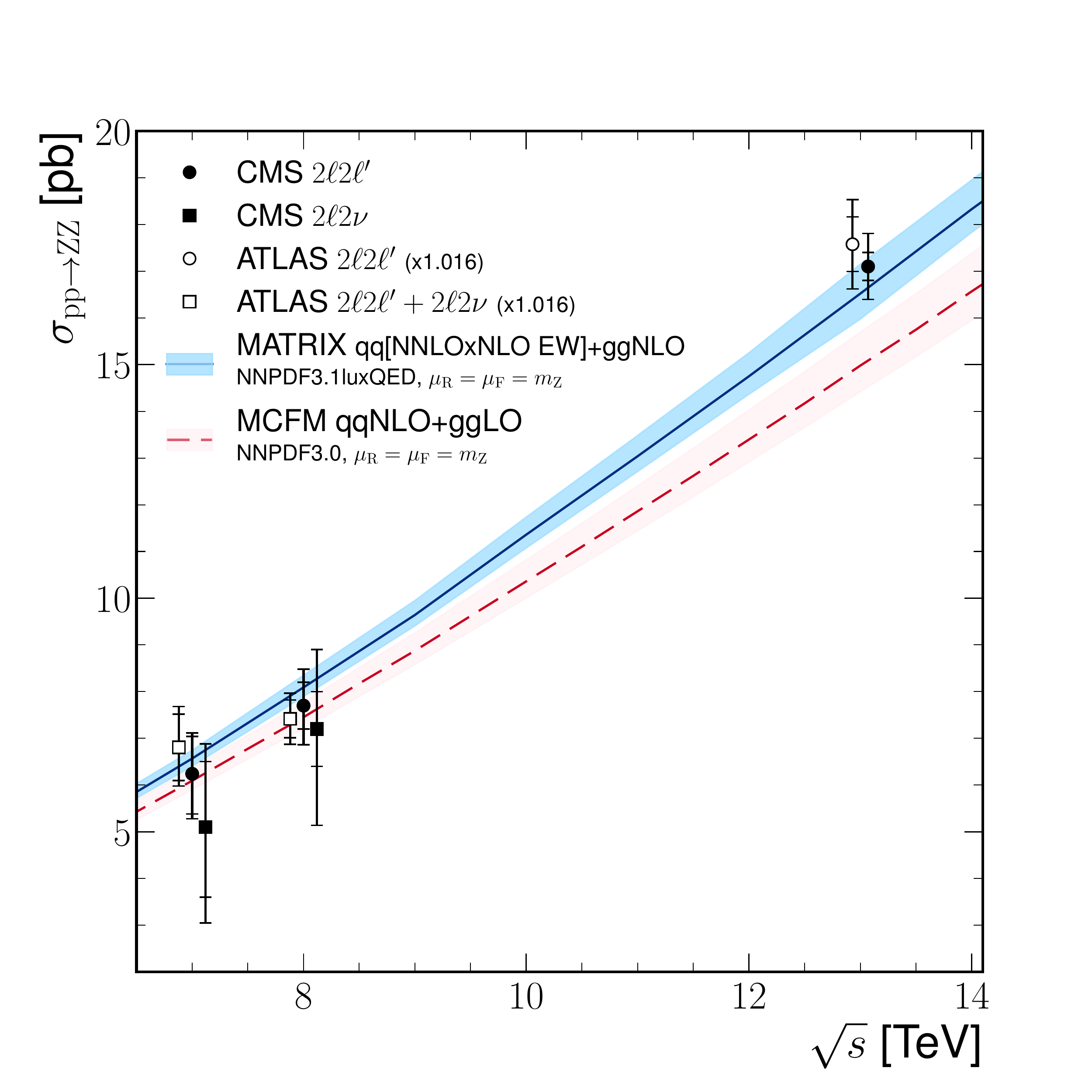} 
\caption{
        The total \ZZ  cross section as a function of the proton-proton
        center-of-mass energy. Results from the CMS~\cite{Chatrchyan:2012sga, CMS:2014xja}
        and ATLAS~\cite{Aad:2012awa,Aad:2015rka,Aaboud:2017rwm} experiments
        are compared to predictions from \matrixMC\ at NNLO in QCD and NLO in EW, 
        and \MCFM at NLO in QCD. The \MCFM prediction also includes gluon-gluon initiated
        production at LO in QCD. The predictions use NNPDF31\_nnlo\_as\_0118\_luxqed and NNPDF3.0 PDF sets, 
        respectively, and fixed factorization and renormalization scales
        $\mu_\mathrm{F} = \mu_\mathrm{R} = m_{\PZ}$. Details of the calculations and uncertainties are given
        in the text. The ATLAS measurements were performed with a $\PZ$ boson
        mass window of 66--116\GeV, instead of 60--120\GeV used by CMS, and are corrected for the resulting 1.6\% difference in acceptance.
        Measurements at the same center-of-mass energy are shifted slightly along the
        horizontal axis for clarity.
}
\label{fig:xsec_vs_sqrts}
\end{figure}

{\tolerance=800
The total $\ZZ$ cross section is shown
in Fig.~\ref{fig:xsec_vs_sqrts} as a function of the \pp
center-of-mass energy. Results from
CMS~\cite{Chatrchyan:2012sga, CMS:2014xja}
and ATLAS~\cite{Aad:2012awa,Aad:2015rka,Aaboud:2017rwm}
are compared to predictions from \matrixMC~v2.0.0\_beta1 and \MCFM.
The uncertainties are statistical (inner bars) and
statistical and systematic combined, as obtained from the fit (outer bars). The band
around the \matrixMC\ predictions reflects scale uncertainties, while
the band around the \MCFM predictions reflects both scale and PDF
uncertainties.
\par}

\section{Differential cross sections}
\label{sec:differentials}

\begin{figure*}[htbp]
\centering
\includegraphics[width=0.45\textwidth]{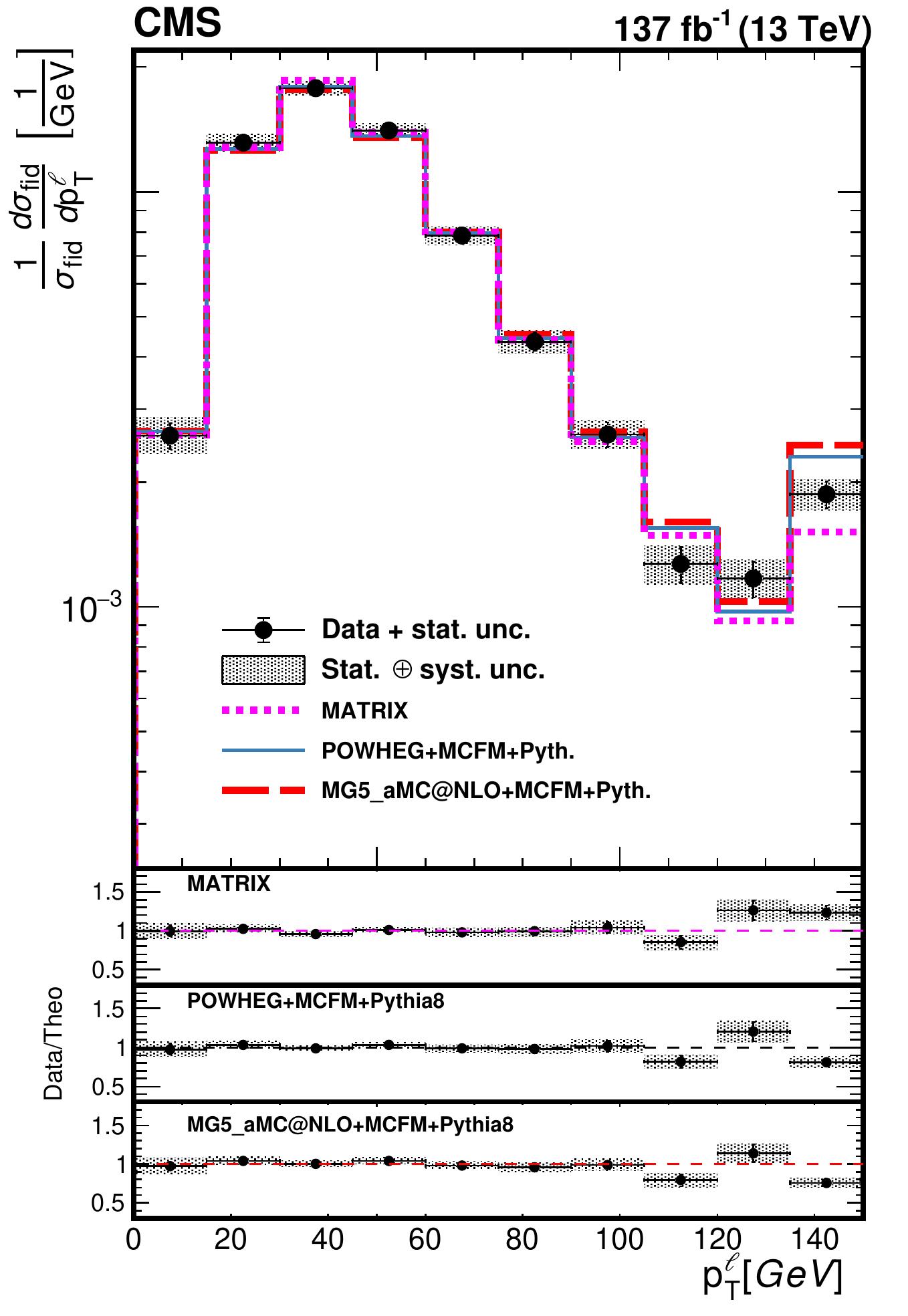}
\includegraphics[width=0.45\textwidth]{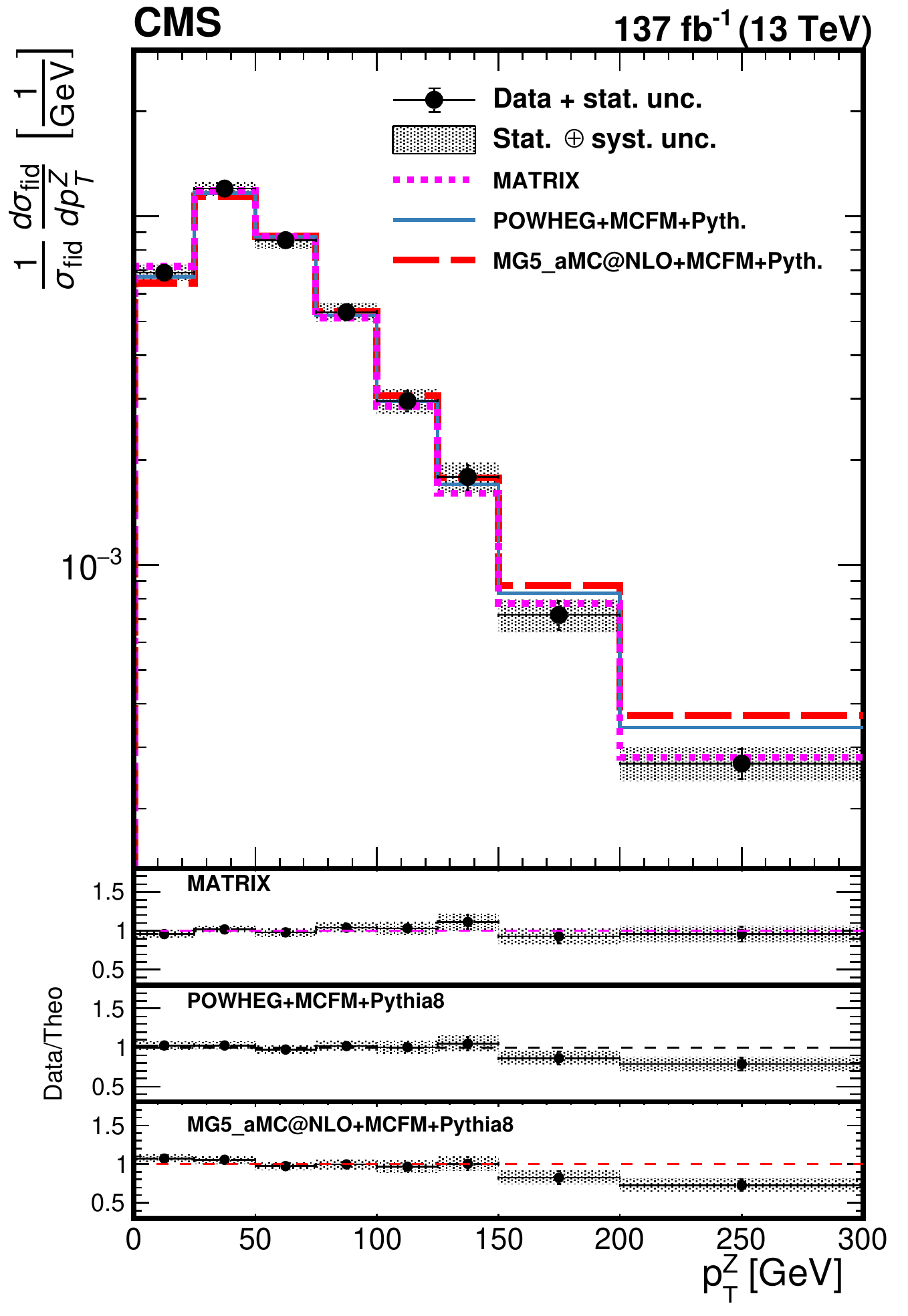}
\caption{Differential cross sections normalized to the fiducial cross section
for the combined 4$\Pe$, 2$\Pe$2$\Pgm$, and 4$\Pgm$ decay channels as a function of
$\pt$ for (left) all leptons, 
(right) all $\PZ$ bosons in the event.
The points represent the unfolded data with error bars showing the statistical uncertainties, the shaded histogram 
the {\POWHEG}+\MCFM \ZZ  predictions,
and the dashed curves correspond to the results of the \matrixMC\ and
{\MGvATNLO}+\MCFM calculations.
The three lower panels represent the ratio of the measured cross section
to the expected distributions from \matrixMC, {\POWHEG}+\MCFM and
{\MGvATNLO}+\MCFM. 
The shaded
areas in all the panels represent the full uncertainties calculated as the quadratic sum
of the statistical and systematic uncertainties, whereas the crosses represent only the statistical
uncertainties.
}
\label{fig:diff1a}
\end{figure*}

\begin{figure*}[htbp]
\centering
\includegraphics[width=0.45\textwidth]{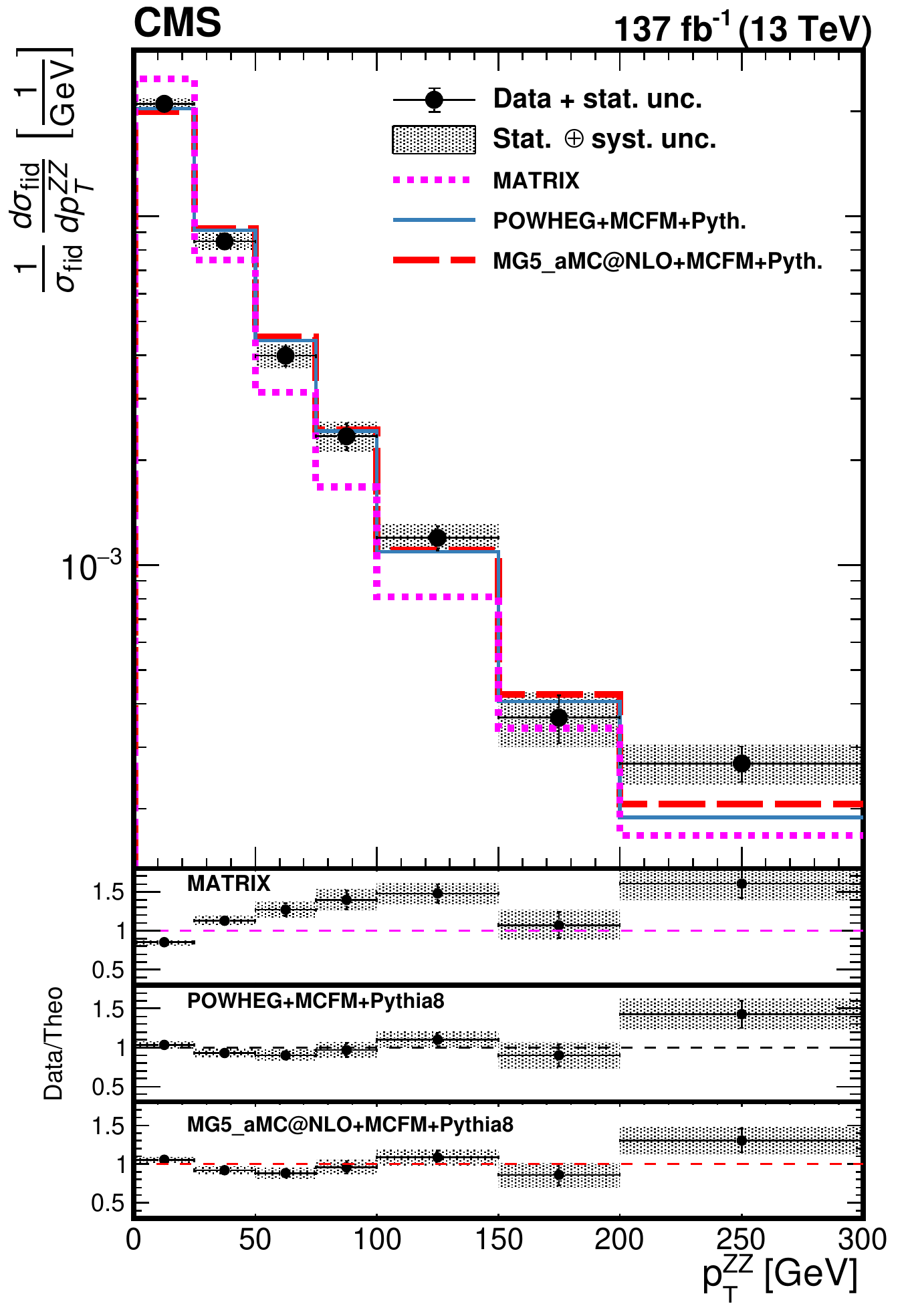}
\includegraphics[width=0.45\textwidth]{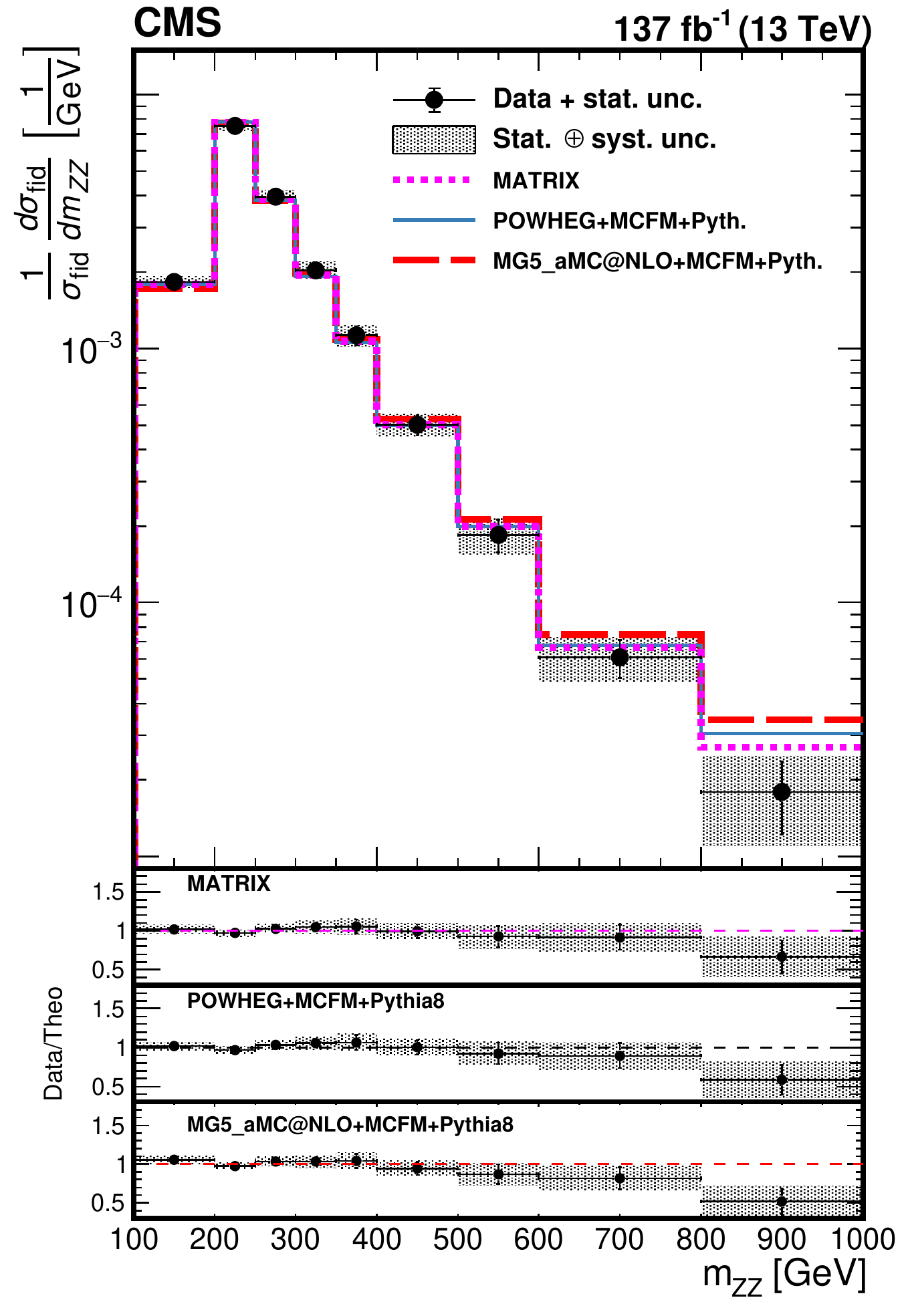}
\caption{Differential cross sections normalized to the fiducial cross section
for the combined 4$\Pe$, 2$\Pe$2$\Pgm$, and 4$\Pgm$ decay channels as a function of
(left)
$\pt$ of the $\ZZ$ system, (right) the
invariant mass of the \ZZ  system.
The points represent the unfolded data with error bars showing the statistical uncertainties, shaded histogram 
the {\POWHEG}+\MCFM \ZZ  predictions,
and the dashed curves correspond to the results of the \matrixMC\ and
{\MGvATNLO}+\MCFM calculations.
The three lower panels represent the ratio of the measured cross section
to the expected distributions from \matrixMC, {\POWHEG}+\MCFM and
{\MGvATNLO}+\MCFM. 
The shaded
areas in all the panels represent the full uncertainties calculated as the quadratic sum
of the statistical and systematic uncertainties, whereas the crosses represent only the statistical
uncertainties. 
}
\label{fig:diff1b}
\end{figure*}

\begin{figure*}[htbp]
\centering
\includegraphics[width=0.45\textwidth]{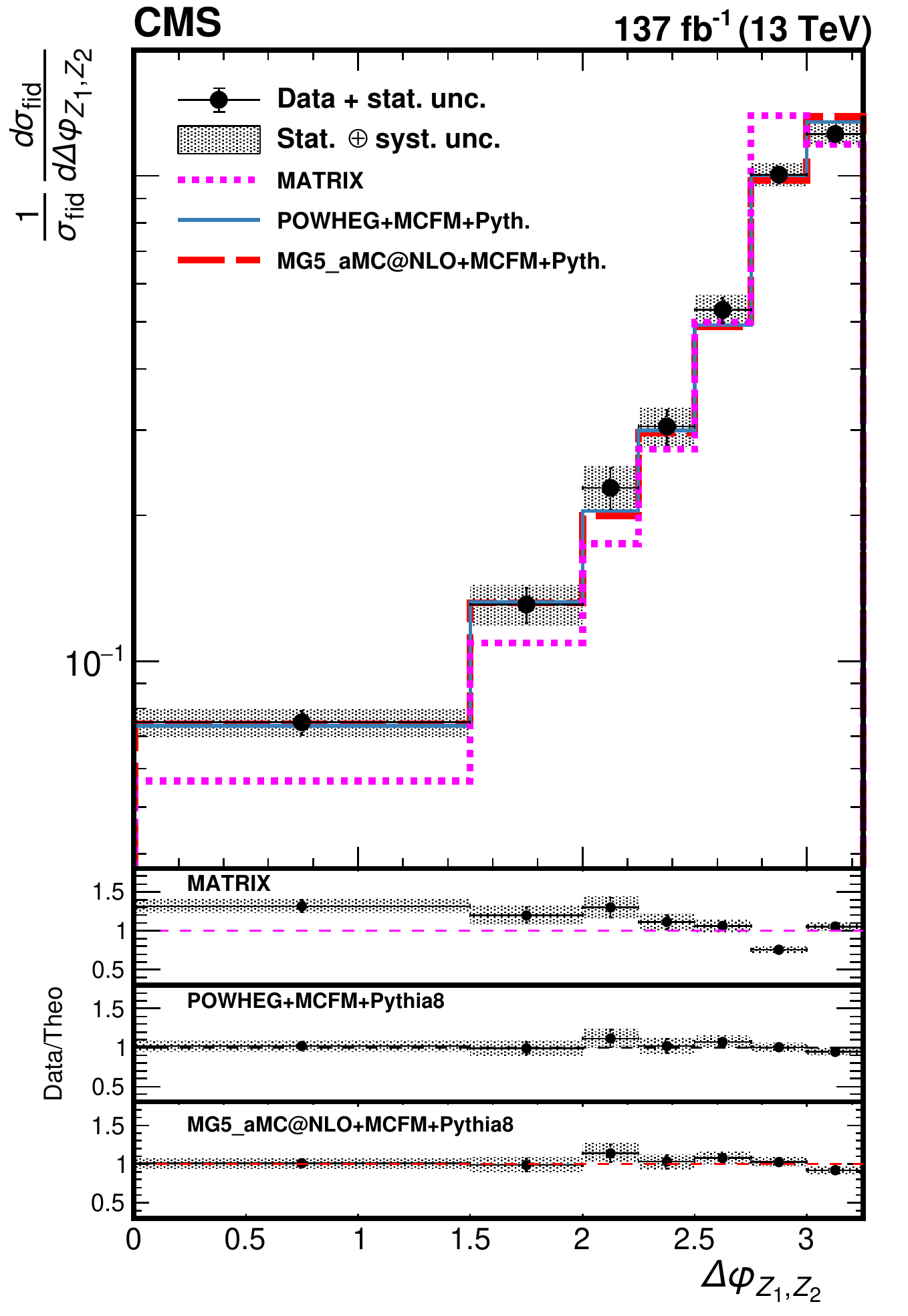}
\includegraphics[width=0.45\textwidth]{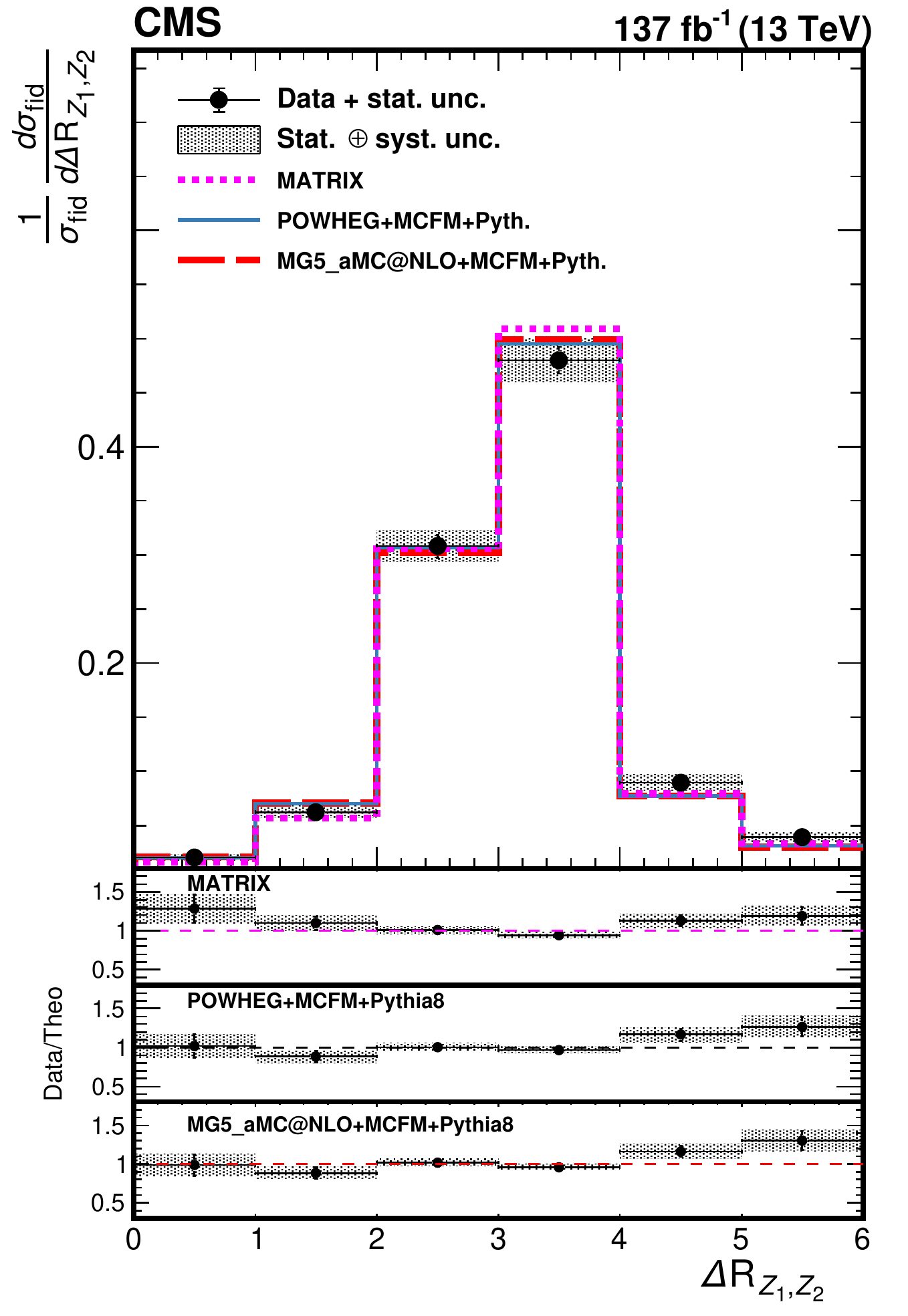}
\caption{
Differential cross sections normalized to the fiducial cross section
for the combined 4$\Pe$, 2$\Pe$2$\Pgm$, and 4$\Pgm$ decay channels as a function of
the azimuthal (left) and $\Delta R$ (right) separation of the two $\PZ$ bosons.
The points represent the unfolded data with error bars showing the statistical uncertainties, the shaded histogram 
the {\POWHEG}+\MCFM \ZZ  predictions,
and the dashed curves correspond to the results of the \matrixMC\ and
{\MGvATNLO}+\MCFM calculations.
The three lower panels represent the ratio of the measured cross section
to the expected distributions from \matrixMC, {\POWHEG}+\MCFM and
{\MGvATNLO}+\MCFM. 
The shaded
areas in all the panels represent the full uncertainties calculated as the quadratic sum
of the statistical and systematic uncertainties, whereas the crosses represent only the statistical
uncertainties.
}
\label{fig:diff2}
\end{figure*}

{\tolerance=800
The differential distributions normalized to the fiducial cross sections are
presented
in Figs.~\ref{fig:diff1a}--\ref{fig:diff2} for the combination of the 4$\Pe$, 2$\Pe$2$\Pgm$, and 4$\Pgm$
decay channels using the whole data sample. The fiducial cross section definition includes $\pt^{\ell}$ and
$\abs{\eta^{\ell}}$ selections on each lepton,
and the 60--120\GeV mass requirement, as described in Section 4.
Figure~\ref{fig:diff1a} shows the differential cross sections in bins of $\pt$ for:
(left) all leptons in the
event, (right) both  $\PZ$ bosons in the event, and in Fig.~\ref{fig:diff1b} (left) for the $\pt$ of the $\ZZ$ system. Figure~\ref{fig:diff1b} (right) shows the normalized
$\rd\sigma/\rd{m_{\ZZ}}$
distribution. 
All \pt and $m_{\ZZ}$ distributions include overflow in the last bin.
Figure~\ref{fig:diff2} shows the angular correlations between
$\PZ$ bosons. 
The data are corrected for background contributions and unfolded for detector effects using
a matrix inversion method without regularization as described in Ref.~\cite{Adye:2011gm},
 and compared with
the theoretical predictions from {\POWHEG}+\MCFM, {\MGvATNLO}+\MCFM and \matrixMC. 
The distributions include both \PZ boson candidates or all four leptons, where applicable, and are normalized
to the numbers of objects in the event and to the fiducial cross section.
The bottom part of each plot shows
the ratio of the measured to the predicted values. The bin sizes are chosen according to
the resolution of the relevant variables, trying also to keep the statistical
uncertainties at a similar level for all the bins.
\par}

{\tolerance=800
The distributions predicted by {\POWHEG}+\MCFM and 
 {\MGvATNLO}+\MCFM agree well with data, except for $m_{\ZZ}$.
This distribution shows a small overestimate in the cross section at high invariant masses. The \matrixMC\ predictions describe this
region better, which can be explained by the presence of the EW corrections in the \matrixMC\ calculations.
 The effect of EW corrections is in detail discussed in Ref.~\cite{Grazzini_2020} and can reach 20--30\% for $m_{\ZZ} = 1 \TeV$.
 On the other hand,
the  \matrixMC predictions show some deviation from the measurements as a function of $\pt^{\ZZ}$ and for
the azimuthal separation between the two \PZ bosons, which is not observed for {\POWHEG}+\MCFM and
 {\MGvATNLO}+\MCFM predictions.
\par}

\section{Limits on anomalous triple gauge couplings}
\label{sec:aTGC}

\begin{figure}[htbp]
\centering
\includegraphics[width=0.95\linewidth]{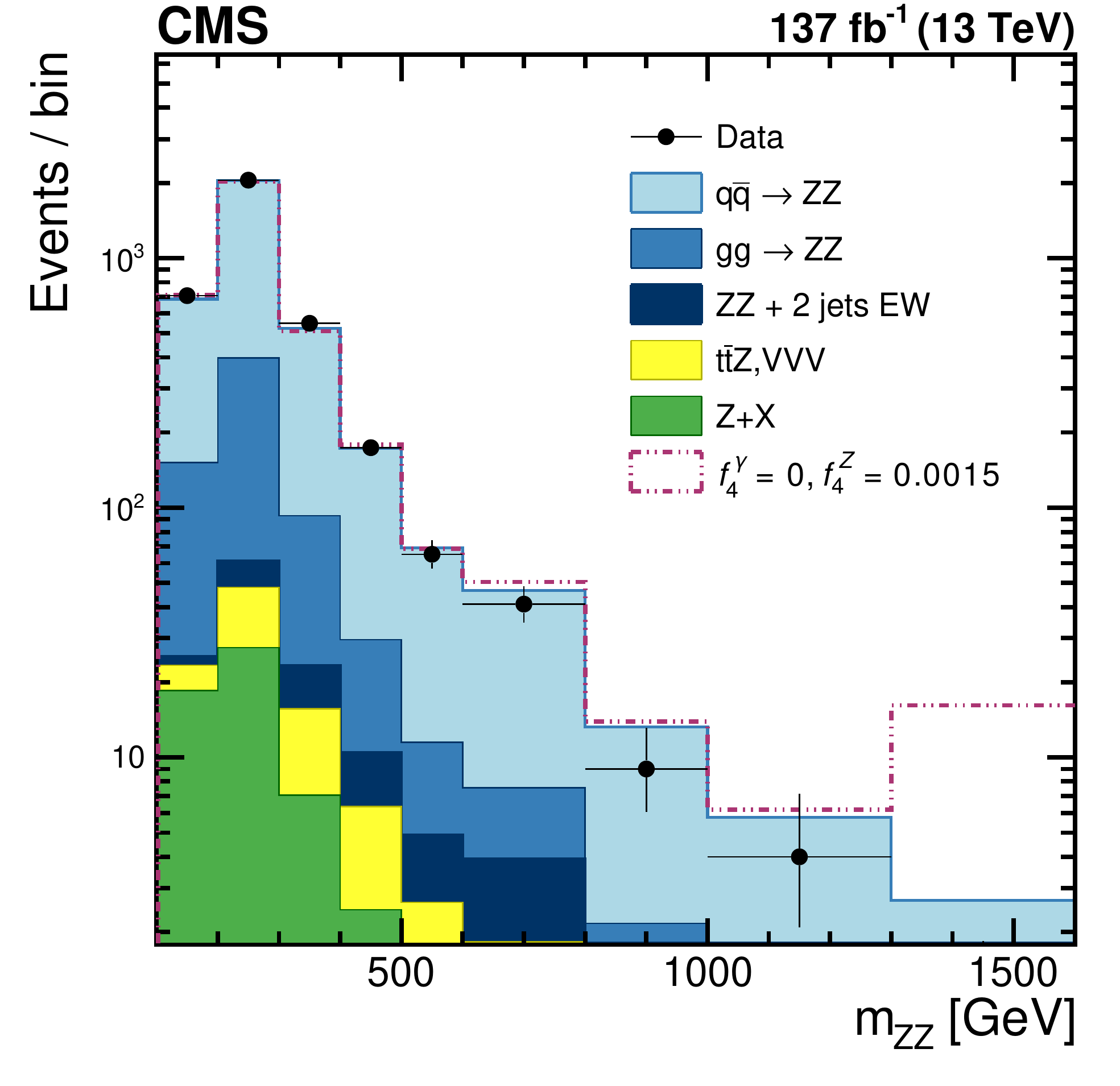}
\caption{
 Distribution of the reconstructed \ZZ  mass for the combined
4$\Pe$, 2$\Pe$2$\Pgm$, and 4$\Pgm$ channels.
Points represent the data with error bars showing the statistical uncertainties, the shaded histograms represent the SM
prediction including signal and irreducible background from simulation, and the
reducible background estimate from data. Dashed histogram represents an example of the aTGC
signal.
The last bin includes 
contribution from all events with  mass above 1300\GeV.
}
\label{figure:sherpa4l}
\end{figure}

The presence of aTGCs
is expected to  increase the event yield at high four-lepton masses.
Figure~\ref{figure:sherpa4l} presents the
distribution of the four-lepton reconstructed mass
for the combined
4$\Pe$, 2$\Pe$2$\Pgm$, and 4$\Pgm$ channels, for the SM and an example of nonzero aTGC value with $f_4^\gamma$=0, and $f_4^\PZ$=0.0015.
Limits on aTGCs are derived from fits to this distribution.
The shaded histograms represent the SM predictions as described in the previous sections and
the dashed curve shows the \SHERPA prediction. The \SHERPA SM predictions are normalized to 
the {\POWHEG}+{\MCFM}  predictions including $K$~factors and agree well with them in shape, as shown
in Fig.~\ref{figure:sherpa4l}. 
As a cross-check of the procedure, the \SHERPA SM distribution was also corrected bin-by-bin to the {\POWHEG}+{\MCFM} distribution, no
difference was observed 
 in the extracted limits.
The presence of aTGC contribution
increases the expected event yields at masses above 1300\GeV. In the fit, described below, 
this region is subdivided into two bins: 1300--2000\GeV and above 2000\GeV. Typically 60--70\% of the aTGC events 
have  masses above 2000\GeV, whereas the expected SM contribution is approximately
2.4 and 0.2 events in the 1300--2000\GeV and above 2000\GeV bins, respectively.

\begin{figure}[htbp]
\centering
\includegraphics[width=0.48\textwidth]{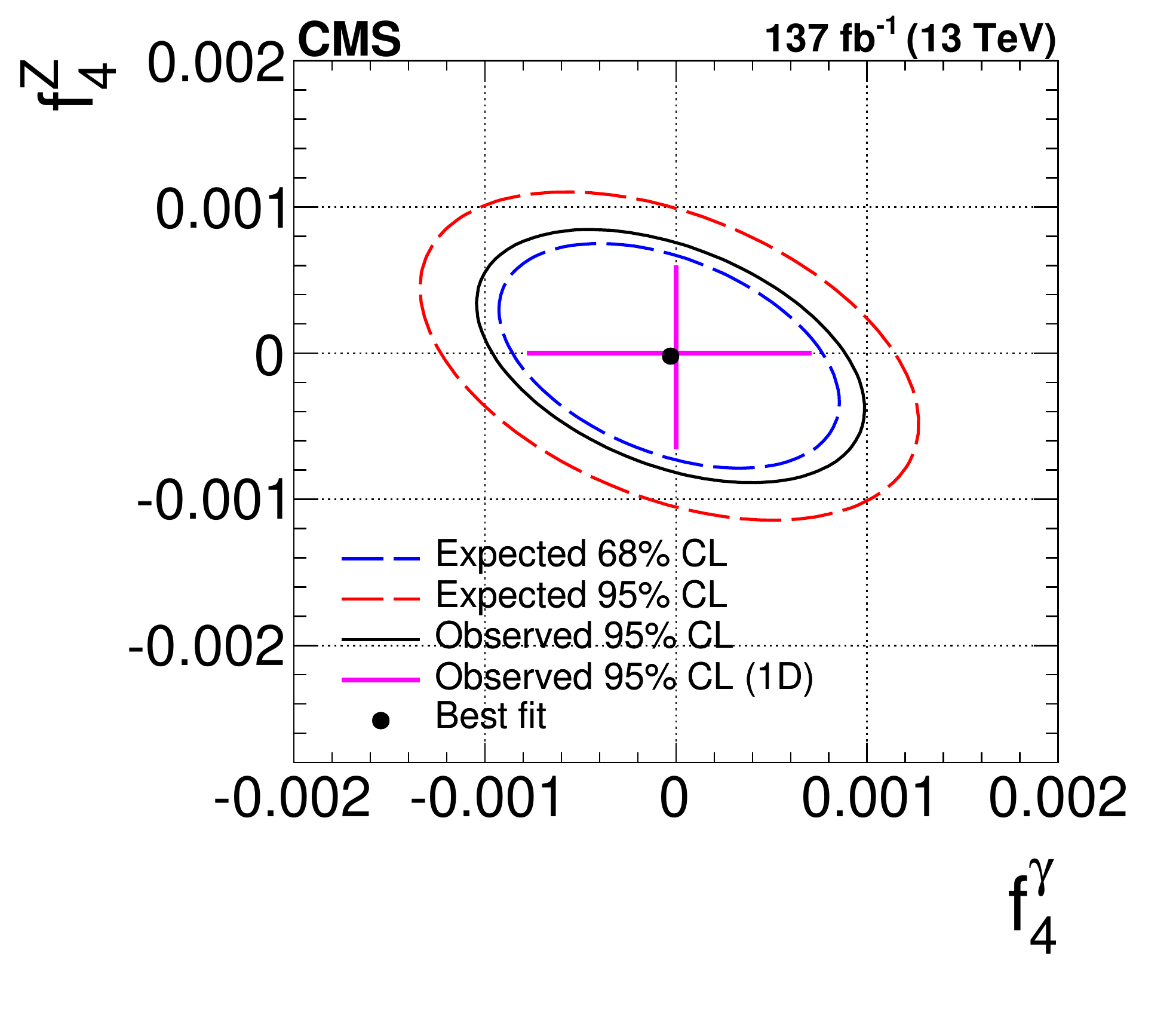}
\includegraphics[width=0.48\textwidth]{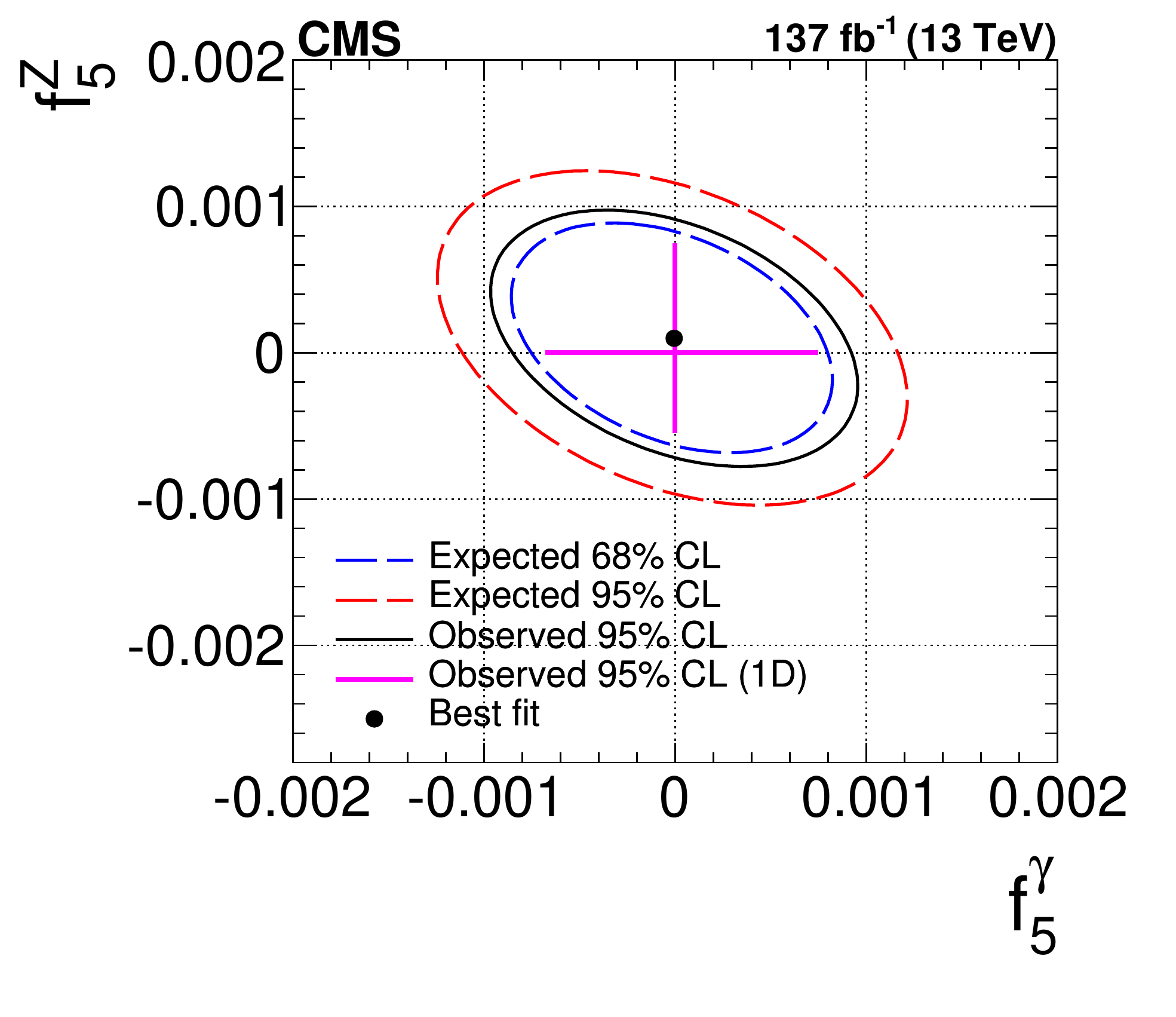}
\caption{Two-dimensional observed (solid) and expected (dashed) contours
 exclusion limits at 95\% \CL, and  at 68 and 95\% \CL, respectively,
 on the $\ZZZ$ and $\ZZ\gamma$ aTGCs.
The plots show the exclusion contours in the
$(f_{4(5)}^\PZ, f_{4(5)}^\gamma)$ parameter planes. Dots show where
the likelihoods reach their maximum. The coupling values outside the contours
are excluded at the corresponding confidence level. The crosses in the middle represent the observed 1D limits.
 No form factor is used. }
\label{figure:aTGC}
\end{figure}

The invariant mass distributions are interpolated from those obtained  from the
\SHERPA simulation for different values of the anomalous couplings in the 
range between 0 and 0.03. For
 each distribution, only one or two couplings are varied while all others 
are set to zero, thus creating a grid of points
in the $(f_{4}^\PZ, f_{4}^\gamma)$ and
$(f_{5}^\PZ, f_{5}^\gamma)$ parameter planes and the corresponding invariant mass
distributions.
In each $m_{\ZZ}$ bin, expected signal values are interpolated between the
two-dimensional grid points using a second-order polynomial, since the cross
section for the signal depends quadratically on the coupling parameters.
A simultaneous fit to the values of aTGCs is performed for all lepton channels, see Ref.~\cite{ATLAS:2011tau} for details.
A profile likelihood method~\cite{Tanabashi:2018oca},
 Wald Gaussian approximation, and Wilks theorem~\cite{Cowan:2010js}  are used to derive
one- (1D) and two-dimensional limits at 68 and 95\% confidence levels (\CL) on each of the aTGC parameters and 
combination of two of them, while all other parameters are set to their SM values.
All systematic uncertainties are included by
varying the number of expected signal and background events within their uncertainties.
An additional 10\% uncertainty is applied on the predictions of the SM and aTGC models
to account for possible differences between model predictions and the interpolation used in the fit.
No form factor~\cite{Baur:2000ae} is used when deriving the limits; 
the results assume that the energy scale of new physics is very high.
The constraints on anomalous couplings are displayed in Fig.~\ref{figure:aTGC}.
The curves indicate
68 and 95\% \CL contours;
the dots indicate where the likelihoods reach
their maximum. Coupling values outside the contours
are excluded at the corresponding \CL.
The crosses in the middle represent the observed 1D limits that are summarized in Table~\ref{table:aTGC}.
The sensitivity is dominated by the statistical uncertainties.

\begin{table}[htbp]
\centering
\topcaption{
  Expected and observed one-dimensional 95\% \CL limits on  aTGC parameters.
  The corresponding constrains on EFT parameters are estimated using the transformation from 
  Ref.~\cite{Degrande:2013kka}.
}
\begin{tabular}{ccc}
\hline
 & Expected 95\% \CL & Observed 95\% \CL  \\
\hline
aTGC parameter & $\times 10^{-4}$ & $\times 10^{-4}$   \\[\cmsTabSkip]
$ f_4^\PZ $ & -8.8  ;  8.3 & -6.6  ;  6.0 \\
$ f_5^\PZ $ & -8.0  ;  9.9 & -5.5  ;  7.5 \\
$ f_4^{\gamma} $ & -9.9  ;  9.5 & -7.8  ;  7.1 \\
$ f_5^{\gamma} $ & -9.2  ;  9.8 & -6.8  ;  7.5 \\[\cmsTabSkip]
EFT parameter & $\TeVns^{-4}$ & $\TeVns^{-4}$   \\[\cmsTabSkip]
$ C_{\mathrm{\tilde{B}}\PW}/\Lambda^4$  & -3.1  ;  3.3 & -2.3  ;  2.5 \\
$ C_{\PW\PW}/\Lambda^4$  & -1.7  ;  1.6 & -1.4  ;  1.2 \\
$ C_{\mathrm{B}\PW}/\Lambda^4$  & -1.8  ;  1.9 & -1.4  ;  1.3 \\
$ C_{\mathrm{BB}}/\Lambda^4$  & -1.6  ;  1.6 & -1.2  ;  1.2 \\
\hline
\end{tabular}
\label{table:aTGC}
\end{table}

Complete one-loop EW
corrections to massive
vector boson pair production~\cite{Bierweiler:2013dja, Baglio:2013toa}
were applied as a cross-check. The EW corrections
to the \ZZ  production cause the \ZZ  mass
spectrum to fall more rapidly at large masses.  In addition, the
overall cross section decreases by about 4\%.
The effect of NLO EW corrections
is estimated by reweighting 
the SM \SHERPA sample as a function of $m_{\ZZ}$
using weights derived from the
calculations described in Ref.~\cite{Bierweiler:2013dja}.
This reweighting improves
the expected limits by about 4--6\%,
 whereas there is no effect on the observed limits. This is expected, since only the 
SM contribution is
 subject to the EW corrections; they are not applied on aTGCs. The limits are driven by the high mass
 tail above 1300 GeV. In this region the aTGC signal is much larger than the SM, and therefore the EW correction
 on the SM part has a very small effect on the predictions of the SM+aTGC model. This correction is much smaller than 
the uncertainty we apply in the fit procedure.

These results
 can be also expressed in terms of EFT  parameters.
The numerical relations between aTGCs and EFT parameters  
are given in Ref.~\cite{Degrande:2013kka}. The expected and measured limits in terms of EFT
are presented in Table~\ref{table:aTGC}.

\section{Summary}

{\tolerance=800
Four-lepton final states have been studied in
proton-proton collisions at $\sqrt{s} = 13\TeV$ with the CMS detector at the
CERN LHC.
The data sample corresponds to an integrated luminosity of 137\fbinv, collected during 2016--2018.   
 The measured $\pp \to \ZZ$ total cross section is
$\sigma_{\text{tot}} ( \pp \to \ZZ ) = 17.4 \pm 0.3 \stat \pm 0.5 \syst \pm 0.4 \thy \pm 0.3  \lum\unit{pb}$,
where the $\PZ$ boson masses are in the range
$60 < m_{\PZ} < 120\GeV$.
The results
agree with the SM predictions, discussed in Section~\ref{sec:xsec}.
The differential cross sections also agree well with
the SM predictions.
Improved limits on
anomalous $\PZ \PZ \PZ$ and $\PZ \PZ \gamma$ triple gauge
couplings
are established.
These are the most stringent limits to date on anomalous
 $\PZ \PZ \PZ$ and $\PZ \PZ \gamma$ triple gauge
 couplings and they improve the previous strictest results from CMS by $\approx$30--40\%.
\par}

\begin{acknowledgments}

  We congratulate our colleagues in the CERN accelerator departments for the excellent performance of the LHC and thank the technical and administrative staffs at CERN and at other CMS institutes for their contributions to the success of the CMS effort. In addition, we gratefully acknowledge the computing centers and personnel of the Worldwide LHC Computing Grid for delivering so effectively the computing infrastructure essential to our analyses. Finally, we acknowledge the enduring support for the construction and operation of the LHC and the CMS detector provided by the following funding agencies: BMBWF and FWF (Austria); FNRS and FWO (Belgium); CNPq, CAPES, FAPERJ, FAPERGS, and FAPESP (Brazil); MES (Bulgaria); CERN; CAS, MoST, and NSFC (China); COLCIENCIAS (Colombia); MSES and CSF (Croatia); RIF (Cyprus); SENESCYT (Ecuador); MoER, ERC IUT, PUT and ERDF (Estonia); Academy of Finland, MEC, and HIP (Finland); CEA and CNRS/IN2P3 (France); BMBF, DFG, and HGF (Germany); GSRT (Greece); NKFIA (Hungary); DAE and DST (India); IPM (Iran); SFI (Ireland); INFN (Italy); MSIP and NRF (Republic of Korea); MES (Latvia); LAS (Lithuania); MOE and UM (Malaysia); BUAP, CINVESTAV, CONACYT, LNS, SEP, and UASLP-FAI (Mexico); MOS (Montenegro); MBIE (New Zealand); PAEC (Pakistan); MSHE and NSC (Poland); FCT (Portugal); JINR (Dubna); MON, RosAtom, RAS, RFBR, and NRC KI (Russia); MESTD (Serbia); SEIDI, CPAN, PCTI, and FEDER (Spain); MOSTR (Sri Lanka); Swiss Funding Agencies (Switzerland); MST (Taipei); ThEPCenter, IPST, STAR, and NSTDA (Thailand); TUBITAK and TAEK (Turkey); NASU (Ukraine); STFC (United Kingdom); DOE and NSF (USA).
   
  \hyphenation{Rachada-pisek} Individuals have received support from the Marie-Curie program and the European Research Council and Horizon 2020 Grant, contract Nos.\ 675440, 752730, and 765710 (European Union); the Leventis Foundation; the A.P.\ Sloan Foundation; the Alexander von Humboldt Foundation; the Belgian Federal Science Policy Office; the Fonds pour la Formation \`a la Recherche dans l'Industrie et dans l'Agriculture (FRIA-Belgium); the Agentschap voor Innovatie door Wetenschap en Technologie (IWT-Belgium); the F.R.S.-FNRS and FWO (Belgium) under the ``Excellence of Science -- EOS" -- be.h project n.\ 30820817; the Beijing Municipal Science \& Technology Commission, No. Z191100007219010; the Ministry of Education, Youth and Sports (MEYS) of the Czech Republic; the Deutsche Forschungsgemeinschaft (DFG) under Germany's Excellence Strategy -- EXC 2121 ``Quantum Universe" -- 390833306; the Lend\"ulet (``Momentum") Program and the J\'anos Bolyai Research Scholarship of the Hungarian Academy of Sciences, the New National Excellence Program \'UNKP, the NKFIA research grants 123842, 123959, 124845, 124850, 125105, 128713, 128786, and 129058 (Hungary); the Council of Science and Industrial Research, India; the HOMING PLUS program of the Foundation for Polish Science, cofinanced from European Union, Regional Development Fund, the Mobility Plus program of the Ministry of Science and Higher Education, the National Science Center (Poland), contracts Harmonia 2014/14/M/ST2/00428, Opus 2014/13/B/ST2/02543, 2014/15/B/ST2/03998, and 2015/19/B/ST2/02861, Sonata-bis 2012/07/E/ST2/01406; the National Priorities Research Program by Qatar National Research Fund; the Ministry of Science and Higher Education, project no. 02.a03.21.0005 (Russia); the Programa Estatal de Fomento de la Investigaci{\'o}n Cient{\'i}fica y T{\'e}cnica de Excelencia Mar\'{\i}a de Maeztu, grant MDM-2015-0509 and the Programa Severo Ochoa del Principado de Asturias; the Thalis and Aristeia programs cofinanced by EU-ESF and the Greek NSRF; the Rachadapisek Sompot Fund for Postdoctoral Fellowship, Chulalongkorn University and the Chulalongkorn Academic into Its 2nd Century Project Advancement Project (Thailand); the Kavli Foundation; the Nvidia Corporation; the SuperMicro Corporation; the Welch Foundation, contract C-1845; and the Weston Havens Foundation (USA).
\end{acknowledgments}
\ifthenelse{\boolean{cms@external}}{\clearpage}{}
\bibliography{auto_generated}

\cleardoublepage \appendix\section{The CMS Collaboration \label{app:collab}}\begin{sloppypar}\hyphenpenalty=5000\widowpenalty=500\clubpenalty=5000\vskip\cmsinstskip
\textbf{Yerevan Physics Institute, Yerevan, Armenia}\\*[0pt]
A.M.~Sirunyan$^{\textrm{\dag}}$, A.~Tumasyan
\vskip\cmsinstskip
\textbf{Institut f\"{u}r Hochenergiephysik, Wien, Austria}\\*[0pt]
W.~Adam, F.~Ambrogi, T.~Bergauer, M.~Dragicevic, J.~Er\"{o}, A.~Escalante~Del~Valle, R.~Fr\"{u}hwirth\cmsAuthorMark{1}, M.~Jeitler\cmsAuthorMark{1}, N.~Krammer, L.~Lechner, D.~Liko, T.~Madlener, I.~Mikulec, F.M.~Pitters, N.~Rad, J.~Schieck\cmsAuthorMark{1}, R.~Sch\"{o}fbeck, M.~Spanring, S.~Templ, W.~Waltenberger, C.-E.~Wulz\cmsAuthorMark{1}, M.~Zarucki
% ...
\vskip\cmsinstskip
\dag: Deceased\\
1:  Also at Vienna University of Technology, Vienna, Austria\\
% ...
\end{sloppypar}
\end{document}